\documentclass[12pt]{article}
\usepackage{amsmath, amsthm}
\usepackage{enumerate}
\usepackage{booktabs}
\usepackage{natbib}
\usepackage{url} 
\usepackage{soul}
\usepackage{amssymb}
\usepackage{comment}
\usepackage{xcolor}
\usepackage{longtable}
\usepackage{booktabs}
\usepackage{cprotect}
\usepackage{color}
\usepackage{algorithm}
\usepackage{rotating}
\usepackage[algo2e,lined,commentsnumbered,ruled]{algorithm2e} 
\usepackage{algpseudocode}
\usepackage[colorlinks=true,linkcolor=blue,citecolor=blue,urlcolor=blue]{hyperref}
\usepackage{amssymb,amsmath}
\usepackage{amsthm}
\usepackage{graphicx}
\usepackage{helvet}
\usepackage[font=footnotesize,labelfont=bf]{caption}
\usepackage{longtable}
\usepackage{mathtools}
\usepackage{xurl}
\usepackage{caption}
\usepackage{subcaption}
\usepackage{pdflscape}
\usepackage{url}
\usepackage{changepage}
\usepackage{multirow}
\usepackage{ulem}
\usepackage{tabularx}
\usepackage{threeparttablex} 
\usepackage{makecell}
\usepackage{tcolorbox}

\usepackage{hyperref}
\hypersetup{
    colorlinks=true,
    citecolor=blue,
    urlcolor=blue,
    hypertexnames=true 
}

\usepackage{graphicx}

\newtheorem{theorem}{Theorem}[section]
\newtheorem{example}[theorem]{Example}

\newcommand{\m}[1]{\mathbf{#1}}
\newcommand{\mb}[1]{\boldsymbol{#1}}

\newcommand{\blind}{0}

\begin{document}

\def\spacingset#1{\renewcommand{\baselinestretch}%
{#1}\small\normalsize} \spacingset{1}


\if0\blind
{
  \title{\bf \textcolor{black}{Failure of Optimal Design Theory? A Case Study in Toxicology Using Sequential Robust Optimal Design Framework}}%
  \author{Elvis Han Cui$^1$, Michael Collins$^2$, Jessica Munson$^2$, Weng Kee Wong$^1$\thanks{
    The authors gratefully acknowledge \textit{please remember to list all relevant funding sources in the unblinded version}}\hspace{.2cm}\\
    1 Department of Biostatistics\\
    2 Department of Environmental Health Sciences \\
    University of California, Los Angeles}
  \maketitle
} \fi

\if1\blind
{
  \bigskip
  \bigskip
  \bigskip
  \begin{center}
    {\LARGE\bf Failure of Optimal Design Theory? A Case Study in Toxicology Using Sequential Robust Optimal Design Framework}
\end{center}
  \medskip
} \fi

\bigskip
\begin{abstract}

This paper presents a quasi-sequential optimal design framework for toxicology experiments, specifically applied to sea urchin embryos. The authors propose a novel approach combining robust optimal design with adaptive, stage-based testing to improve efficiency in toxicological studies, particularly where traditional uniform designs fall short. The methodology uses statistical models to refine dose levels across experimental phases, aiming for increased precision while reducing costs and complexity. Key components include selecting an initial design, iterative dose optimization based on preliminary results, and assessing various model fits to ensure robust, data-driven adjustments. Through case studies, we demonstrate improved statistical efficiency and adaptability in toxicology, with potential applications in other experimental domains.
\end{abstract}


\noindent%
{\it Keywords:}  Robust optimal design, Toxicology, Sequential design, Dose-response, Sea urchins.
\vfill

\newpage
\spacingset{1.45} 

\section{Motivation and Introduction}
\label{sec:intro}

\subsection{Importance of the Paper}
As statisticians, we should always ask ``what others (say, toxicologists) need", instead of saying ``what we have in our beautiful theory". \textbf{This work represents a groundbreaking collaboration between statisticians and toxicologists, addressing the urgent need to reduce costs in toxicology experiments using sea urchins.} Unlike previous studies in sequential optimal design, which are largely theoretical or simulation-based, our research directly engages with practical challenges faced by toxicologists. \textbf{Surprisingly, our findings reveal that parameter estimates derived from optimal designs often fail to outperform those obtained from conventional uniform designs in practice.}

This apparent discrepancy arises not from the failure of optimal design theory but from a multitude of real-world factors. These include (1) practical constraints, such as adherence to Food and Drug Administration (FDA) guidelines and budgetary limitations \citep{fda_animal_rule, buckley2020drug}; (2) the influence of genetic shifts in experimental organisms, which complicate the transferability of theoretical designs across datasets; and (3) the inherent challenges of implementing complex designs in dynamic experimental settings. These factors highlight the critical gap between theoretical advancements and practical applications in toxicological studies.

To bridge this gap, we propose several remedies:
\begin{enumerate}
    \item \textbf{For statisticians}: Design frameworks must incorporate robust features like control groups and endpoints (e.g., lethal doses) to align with biological and practical considerations.
    \item \textbf{For toxicologists}: Greater control over genetic shifts in experimental organisms can improve the consistency and applicability of optimal designs.
    \item \textbf{For researchers and practitioners}: Developing accessible, user-friendly tools is essential to implement theoretical designs effectively in real-world toxicology.
\end{enumerate}

While the future of optimal design in practice remains uncertain, our work highlights the need for continuous adaptation and collaboration. By addressing these challenges, we aim to strengthen the connection between theoretical statistics and experimental toxicology, paving the way for more impactful research and applications.


\subsection{Introduction}
In this era of rapid advancement in statistical methods and computational technology, the time has come to revisit our approach to toxicology experiments. The once-dominant uniform design, which takes an equal number of observations at each of the equally spaced doses  was conceived in an age before the dawn of computational simulations, has \st{now reveals its} limitations, and it is susceptible to empiricism and unable to incorporate improved statistical methodology from the field of optimal designs \citep{tse2014selected}. \textcolor{black}{With} at our disposal now, the future calls for multiple parallel iterative approaches, where experimentation moves in concert with technology. This shift is not a choice, but an inevitability. The march of progress is spiral—driven by constant iterations, failures, and self-renewal \citep{marx2000karl}.

 The field of experimental toxicology is a fascinating and dynamic arena within the scientific landscape. Unlike many traditional disciplines where experiments are conducted under controlled, predictable conditions, toxicology embraces the inherent unpredictability of biological responses. In fields such as chemistry or physics, experiments often follow meticulously designed trajectories, with every variable tightly controlled to minimize surprises. Toxicologists, on the other hand, explore the intricate interactions between substances and biological systems, frequently uncovering unexpected adverse effects. This unpredictability is precisely what makes toxicology both challenging and captivating—every experiment has the potential to reveal something new and unforeseen about the way our bodies interact with the world around us \citep{schoen1996statistical, costa2010design}.

In contemporary toxicology laboratories, researchers frequently adopt a stepwise experimental approach, where the outcomes of each testing phase guide subsequent experiments \citep{collins2022model}. This method, while rooted in decades of practical wisdom, often lacks the efficiency and precision that a more structured experimental design could provide. Enter the world of optimal experimental design—a transformative approach that could revolutionize toxicology. Imagine a process where every experiment is not just an isolated trial, but a part of a strategic, interconnected sequence that builds upon the last. With optimal design, toxicologists can use the data they gather to refine their approach in real-time, saving time, reducing costs, and accelerating the discovery process with precision and accuracy \citep{de2023methodology}.

Historically, toxicologists have relied on straightforward experimental designs, such as evenly spaced dose levels or uniform designs \citep{casarett2008casarett, fedorov2013optimal}. While simple to implement, these approaches often lead to inefficiencies—extended timelines, unnecessary costs, and suboptimal use of data. Moreover, the complexity of toxicological systems, including dose-response nonlinearity, low-dose effects, and phenomena like hormesis \citep{calabrese2003hormesis}, necessitates experimental strategies that adapt to emerging insights. This demands a move away from static, empirical methods to a more dynamic, data-driven approach that integrates modern statistical tools \citep{dragalin2008adaptive,gertsch2024interactive}.

The secret to designing these efficient experiments lies in their flexibility and precision \citep{holland2015optimal}. By clearly defining the goals of a toxicology study—whether it’s identifying a toxic threshold, studying low-dose effects, or detecting complex phenomena like hormesis—researchers can leverage statistical models to guide their experimental designs. This approach maximizes the accuracy of estimates while streamlining the experimental process. Even though optimal design theory wasn’t initially created with sequential experiments in mind, it offers a rich toolbox that can be adapted to meet the evolving needs of toxicology \citep{cui2024optimal}.

In this paper, we present an innovative framework that marries the precision of optimal design theory with the adaptive, step-by-step nature of toxicological testing \citep{koutra2021optimal}. Our method uses data from each experimental phase to fine-tune subsequent steps, ensuring that each iteration propels researchers closer to their objectives. By applying optimal design criteria, toxicologists can select dose levels that maximize statistical precision, all while reducing both costs and labor. Here’s a glimpse into our step-by-step approach:
\begin{itemize}
    \item \textbf{Initial Experiment Design}: Based on their expertise, the toxicologist selects an initial set of doses and conducts the first round of experiments.
    \item \textbf{Modeling the Data}: After analyzing the initial data, the toxicologist identifies the best-fitting statistical model, which will guide the design of the next batch of experiments.
    \item \textbf{Optimal Dose Selection}: Using the fitted model, optimal design criteria can help choose the next set of doses. For example, a D-optimal design focuses on estimating model parameters with maximum accuracy, while a G-optimal design minimizes the variance across the entire dose-response surface. Alternatively, an integrated criterion can emphasize different parts of the response curve, depending on the specific goals of the study. These designs often require fewer dose levels than traditional approaches, making the next round of experiments more efficient.
    \item \textbf{Evaluating and Adjusting}: After the second batch of experiments, the data are analyzed again. The model is either confirmed through graphical or statistical checks (e.g., a lack-of-fit test) or adjusted to reflect new insights. If the model remains consistent with earlier findings, the efficiency of the current design is compared to that of the optimal design.
    \item \textbf{Refining the Design}: If the current design is found to be less efficient, additional dose levels may be introduced to enhance precision. The process is repeated as necessary.
\end{itemize}

The following sections demonstrate how our method can be applied to real-world toxicology experiments, showcasing significant improvements in efficiency and cost-effectiveness. By embracing this streamlined, data-driven approach, toxicologists can transform the way they approach experimental design, moving from initial hypotheses to conclusive findings with greater speed and precision. The rest of the paper is organized as follows: Section~\ref{sec:review} provides a comprehensive review of existing sequential optimal design methodologies and highlights their limitations in toxicological applications. Section~\ref{sec:meth} introduces our proposed quasi-sequential robust optimal design framework, detailing its theoretical foundation and implementation. Section~\ref{ref:case_studies_design} applies this framework to real-world toxicology experiments, showcasing its effectiveness in improving dose-response modeling and experimental efficiency. Section~\ref{sec:simu} provides simulation studies based on bivariate probit model. Finally, Section 5 discusses the implications of our findings, potential limitations, and directions for future research.

\section{Literature Review: Sequential Optimal Design and Applications}\label{sec:review}

Sequential optimal design has been a cornerstone in experimental research, allowing for iterative refinement of experimental parameters based on accumulated data. This approach has found applications across various fields, including toxicology, clinical trials, and epidemiology.

\textbf{Theoretical Foundations}: The concept of sequential design traces its roots to early works in optimal experimental design, such as those by \cite{silvey2013optimal} and \cite{fedorov1971design}. Sequential designs aim to enhance efficiency by adapting experimental conditions dynamically. \cite{park1998sequential} introduced a nonparametric sequential approach for estimating response curves, emphasizing the advantage of iterative adaptation in achieving greater precision with fewer samples. Similarly, \cite{hughes2000efficient} utilized compound D-optimality in adaptive group testing to estimate rare trait prevalences efficiently, underscoring the method's applicability to nonlinear and heteroscedastic models. \cite{lane2020adaptive} developed a theoretical framework that utilizes observed Fisher information as a priori.

\textbf{Applications in Dose-Response Studies}: One prominent application of sequential optimal design is in dose-response studies. \cite{dragalin2008two} proposed a two-stage design for dose-finding in clinical trials, incorporating both efficacy and safety considerations through a bivariate probit model. This methodology demonstrated how sequential designs could address ethical and practical constraints in clinical research by optimizing dose allocations iteratively. \cite{stacey2007adaptive} extended this idea using a Bayesian framework, showcasing how adaptive designs could improve the precision of dose-response models while reducing the risk to participants.  

\textbf{Real-World Implementations}: Real-world implementations of sequential designs have highlighted their versatility. For instance, \cite{wright2006optimal} discussed sequential methods in toxicology, where adaptive designs facilitated the efficient identification of toxic thresholds. In biomedical research, \cite{qiu2023nature} employed metaheuristic algorithms like particle swarm optimization to develop optimal designs for continuation-ratio models, addressing complex dose-response relationships in early-phase clinical trials. Similar algorithms are booming in optimal design and other fields of statistics \citep{cui2024applications}.

\textbf{Challenges and Advancements}: Despite their advantages, sequential designs face challenges, including dependence on prior information and computational complexity. Recent advancements aim to mitigate these issues. \cite{de2023methodology} introduced D-augmentation techniques to enhance design flexibility, enabling better model discrimination and adequacy checks. \cite{park1998sequential} highlighted the importance of robust algorithms that adapt to uncertain initial conditions, ensuring reliability in applications with limited prior data.


\section{Methodology}
\label{sec:meth}

In this section, we detail the proposed sequential robust optimal design framework, which builds on existing methodologies such as \citet{wang2013two}. We also outline the mathematical framework, the sequential design process across two stages, and discuss an augmented design approach to account for practical requirements. Finally, we demonstrate the applicability of this framework to proportional odds models commonly used in toxicological studies.

\subsection{Notations}
The sequential robust optimal design aims to iteratively refine experimental designs across multiple stages to achieve greater efficiency and precision. The core idea is to leverage information from earlier stages of the experiment to optimize the design of subsequent stages, ensuring robustness to model uncertainties.

Let $[x_L, x_U]$ denote the design space, where $x_L$ and $x_U$ represent the lower and upper bounds of the experimental range. An initial proportion of samples, $\alpha$, is allocated to Stage I, while the remaining $(1-\alpha)$ samples are reserved for Stage II. The optimality criterion, such as D-optimality, c-optimality, or dual-optimality, guides the design process by optimizing the Fisher information matrix across stages.

\subsection{Proposed Sequential Robust Optimal Design Scheme}\label{sec:proposed_seq_robust_design}

In the following, we propose a sequential robust optimal design scheme which is a modification of \cite{wang2013two}. We provide a comparison between two methods in Table~\ref{tab:comparison}.

\begin{itemize}
    \item \textbf{Inputs}: 
    \begin{itemize}
        \item $[x_L,x_U]$, the design space;
        \item $\alpha$, the proportion of samples performed at Stage I (so $(1-\alpha)$ is the proportion of samples performed at Stage II).
        \item Optimality criterion (e.g., D-optimality, c-optimality or dual-optimality).
    \end{itemize}
    \item \textbf{Stage I}:
    \begin{itemize}
        \item Perform the initial design which could be evenly spaced on the original or log-scale or be designed using toxicologist's expert knowledge;
        \item Collect a total of $N_1$ samples;
        \item Repeat the procedure $K$ times.
        \item Choose a regression model that fits the first set of data well.
    \end{itemize}
    \item \textbf{Stage II}:
    \begin{itemize}
        \item Utilize the information from Stage I to perform Stage II design for the remaining $(1-\alpha)$ experiments:
        \begin{itemize}
            \item Based on the $K$ sets of data and the regression model that we have chosen, we estimate $K$ sets of nominal values.
            \item Based on the optimality criteria we have chosen, we compute the robust optimal design that optimizes the criterion function where the Fisher information matrix is the combination of the first stage design and the to-be-determined second stage design. It is robust in the sense that we utilizes $K$ different sets of nominal values.
            
        \end{itemize}
        \item Perform the statistical modeling and inference using the obtained Stage II design to get the fitted model.
    \end{itemize}
    \item \textbf{Outputs}: The fitted models and the quantity that we are interested in (e.g., parameter estimates, ED50, oral radialization 50, etc.).
\end{itemize}

\subsection{Augmented Optimal Design}
\label{subsec:augmented}
To address practical considerations in toxicological studies, such as the inclusion of control groups, we propose an augmented design approach. The augmented design incorporates additional dose levels to account for specific experimental requirements.

\paragraph{Incorporating a Control Group}
The design can include a zero-dose level to estimate baseline responses, as follows:
\[
\xi = \begin{pmatrix}
0 & x_1 & \cdots & x_n \\
p & (1-p)p_1 & \cdots & (1-p)p_n
\end{pmatrix},
\]
where $p$ is the weight assigned to the control group ($x = 0$) and $p_i$ are the weights for other dose levels. Following \citet{gollapudi2013quantitative}, a common choice is $p = 1/(n+1)$.

\paragraph{Incorporating High Dose Levels}
To capture endpoints such as lethal effects, the design can include an additional high-dose level $x^*$:
\[
\xi = \begin{pmatrix}
0 & x_1 & \cdots & x_n & x^* \\
\alpha_1 & (1-\alpha)p_1 & \cdots & (1-\alpha)p_n & \alpha_2
\end{pmatrix},
\]
where $\alpha_1 + \alpha_2 = \alpha$ and $x^*$ ensures the inclusion of high-dose observations.

\subsection{Application to Proportional Odds Models}
The proposed framework is particularly suitable for proportional odds models, widely used in toxicological studies to analyze ordinal responses. These models relate the cumulative probability of a response to dose levels through a logistic function:
\[
\log \frac{P(Y \leq k \mid x)}{P(Y > k \mid x)} = \beta_0 + \beta_1 x,
\]
where $Y$ denotes the response category, $x$ is the dose level, and $\beta_0, \beta_1$ are model parameters.

\paragraph{Stage I Design for Proportional Odds Models}
In Stage I, dose levels are selected to ensure sufficient coverage of the response range, allowing for accurate parameter estimation. Data from this stage provide initial estimates of $\beta_0$ and $\beta_1$.

\paragraph{Stage II Design for Proportional Odds Models}
Stage II incorporates the estimated parameters into the optimal design criteria. The Fisher information matrix for the proportional odds model is used to determine dose levels that maximize parameter precision while maintaining robustness to uncertainty.

\paragraph{Outputs}
The final model provides parameter estimates and key toxicological endpoints, such as the effective dose (ED50) and thresholds for higher response categories.

\begin{table}[h!]
\centering
\caption{Comparison of Wang et al. (2013) and the Proposed Scheme}
\label{tab:comparison}
\begin{tabular}{l|p{5cm}p{5cm}}
\hline\hline
\textbf{Aspect} & \textbf{Wang et al. (2013)} & \textbf{Proposed Scheme} \\ \hline
\textbf{Framework} & Two-stage design for dose-response modeling. & Sequential robust design adaptable to various contexts. \\ \hline
\textbf{Stage I Design} & Evenly spaced doses. & Flexible: uniform, log-scale, or expert-driven. \\ \hline
\textbf{Stage II Design} & Bootstrap-based optimization. & Robust criteria using Fisher information. \\ \hline
\textbf{Variance Heterogeneity} & Explicitly modeled. & Similar but more flexible. \\ \hline
\textbf{Efficiency} & High computational cost due to bootstrapping. & Faster with analytical optimization. \\ \hline
\textbf{Practicality} & Limited to basic designs. & Supports control groups and extreme doses (augmented design). \\ \hline\hline
\end{tabular}
\end{table}

\subsection{Optimizer}
The Particle Swarm Optimizer (PSO) is the main optimizer used in this paper and it is widely available online \citep{miranda2018pyswarms, riza2019metaheuristicopt}.

\section{Case Study: Toxicology Experiments}\label{ref:case_studies_design}
In this section, we explore the application of our proposed quasi-sequential robust optimal design framework to real-world toxicology experiments using sea urchin embryos. Sea urchin embryos are an important biological model for understanding dose-response relationships due to their sensitivity to environmental toxins and their relevance to broader ecological and human health concerns. These experiments aim to identify critical dose levels, such as the effective dose for 50\% response (ED50) and lethal dose for 50\% mortality (LD50), which are fundamental metrics for assessing toxicity \citep{calabrese2003hormesis, gollapudi2013quantitative}.

The challenges of these studies arise from high biological variability due to genetic heterogeneity and environmental influences. Conventional static experimental designs, such as uniform or evenly spaced dose levels, often fail to capture complex dose-response relationships or adapt to emerging patterns in the data. These limitations result in inefficiencies, including increased costs, wasted resources, and reduced precision in parameter estimates \citep{ritz2015dose, hartung2009toxicology}.

Our proposed framework, leveraging Particle Swarm Optimization (PSO) and robust design principles, addresses these issues by iteratively refining the experimental design across two stages. This approach balances efficiency and flexibility, enabling the exploration of critical dose-response regions while adapting to uncertainties in biological systems. In this case study, we demonstrate the framework's efficacy using experimental data from sea urchin toxicology studies. 

Section~\ref{subsec:data_analysis} details the experimental data collected and its structure, highlighting the variability across different timelines and genetic groups. Section~\ref{subsec:model_fitting} discusses the modeling approaches and diagnostic measures used to evaluate the data. Section~\ref{sec:locally_optimal} and \ref{sec:two_stage_robust} present the optimal designs generated using our framework, comparing them with conventional methods. Finally, Section~\ref{sec:equi_new} provides a theoretical condition that ensures the generated designs are indeed optimal.
\subsection{Description of Four Datasets}
\label{subsec:data_analysis}
\subsubsection{The First Dataset}
The first dataset was collected by Dr. Collins between June 23rd and August 19th, 2022, over nine days of experiments. The dose levels and design weights were determined solely by the toxicologist’s expertise. Table~\ref{table:2} illustrates the typical data structure, capturing essential experimental variables such as dose levels, embryo response categories, and durations.

\begin{table}[h!]
\centering
\begin{tabular}{||c c c c c c c c||} 
 \hline
 date	&dose&	duration	&observed&	normal&	radial&	0 spicules &	dead/delayed \\ [0.5ex] 
 \hline\hline
  2022.6.23 & 0 & 1-24h & 108 & 107 & 0 & 0 & 1 \\
 $\cdots$ & $\cdots$ & $\cdots$ & $\cdots$ & $\cdots$ & $\cdots$ & $\cdots$ & $\cdots$  \\ [1ex] 
 \hline
\end{tabular}
\caption{A typical sea urchin data structure.}
\label{table:2}
\end{table}
\begin{itemize}
    \item Date: the date of the row being recorded.
    \item Dose: dose level with unit $m g/cm^{3}$, the lowest dose level is 0.
    \item Duration: the total experimental time of the sea urchin embryo; 1-24h means that embryo has spent 24 hours in the solution.
    \item Observed: total number of observed sea urchin embryos.
    \item Normal: number of observed normal embryos.
    \item Radial: Number of embryo that has radialization.
    \item 0 spicules and dead/delayed: Together with other unrecorded abnormal status, these categories record the number of observed abnormal sea urchin embryos.
\end{itemize}

\subsubsection{The Second, Third and Fourth Datasets}

The second, third, and fourth datasets were collected based on the framework developed in Section~\ref{sec:meth}, utilizing optimal design theory and PSO. These datasets represent the second-stage designs in our study. Specifically, the second dataset was collected in December 2022, while the third and fourth datasets were gathered between January 2024 and April 2024. Detailed descriptions of the design processes for these datasets, along with their associated statistical analyses (including the first dataset), are provided in Section~\ref{subsec:optimization_results}.

\subsection{Sequential Robust Optimal Designs}
\label{subsec:subsec:optimal_designs}
\subsubsection{Analysis and Model Selection Based on the First Dataset}\label{subsec:model_fitting}
\begin{figure}[!htbp]
\centering
 \subfloat[]{\includegraphics[width=0.32
 \textwidth]{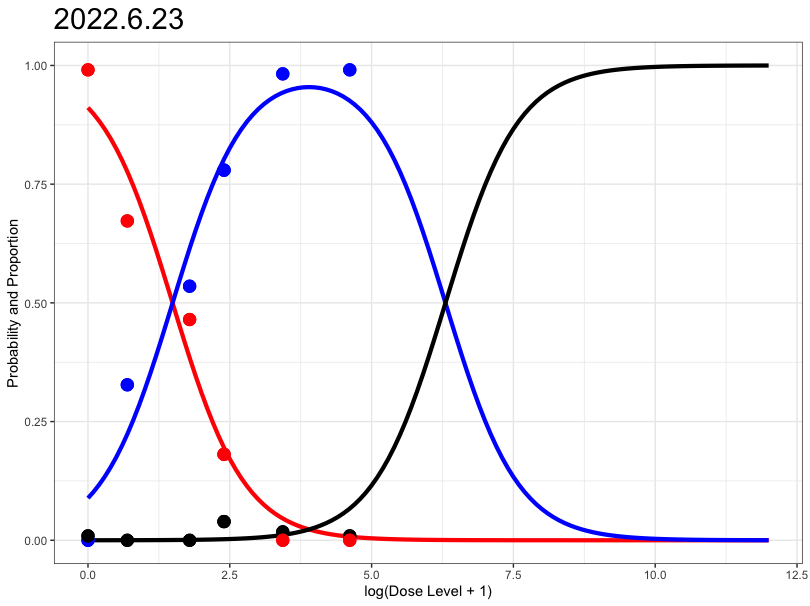}}
\subfloat[]{\includegraphics[width=0.32\textwidth]{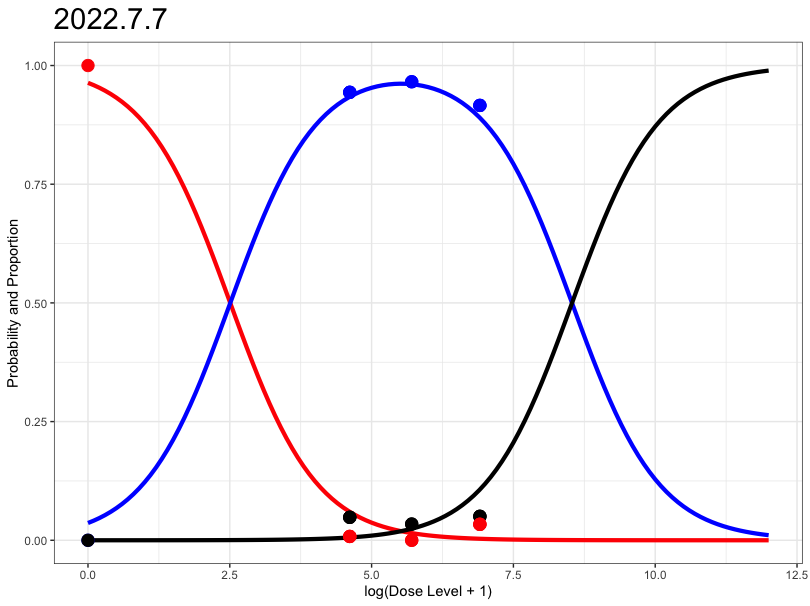}} 
 \subfloat[]{\includegraphics[width=0.32
 \textwidth]{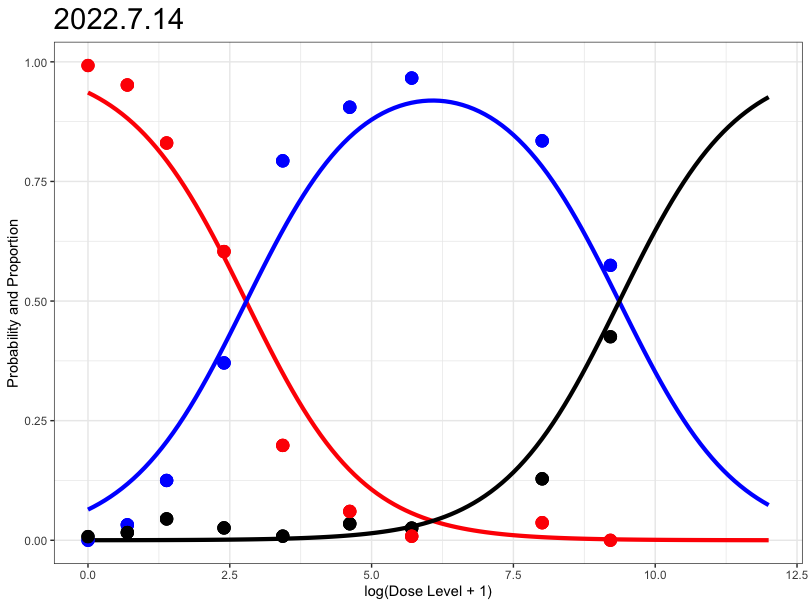}}\\
\subfloat[]{\includegraphics[width=0.32\textwidth]{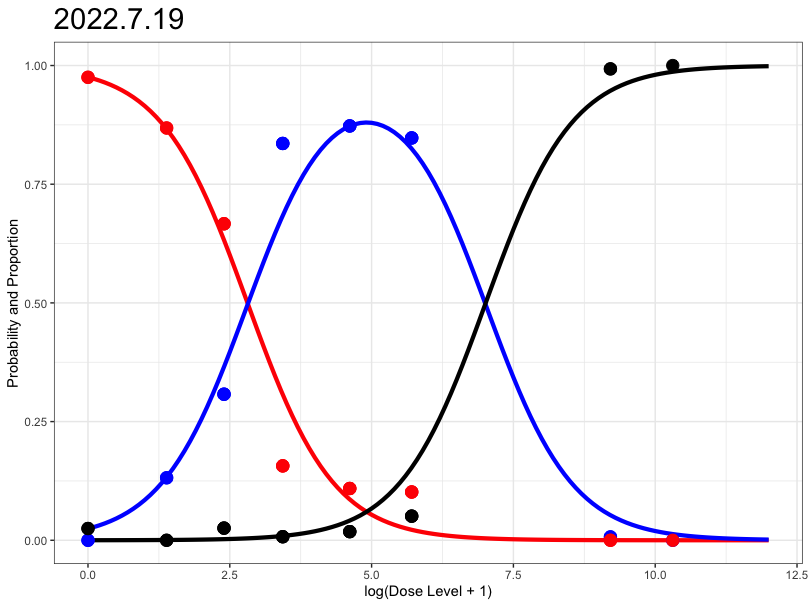}}  
 \subfloat[]{\includegraphics[width=0.32
 \textwidth]{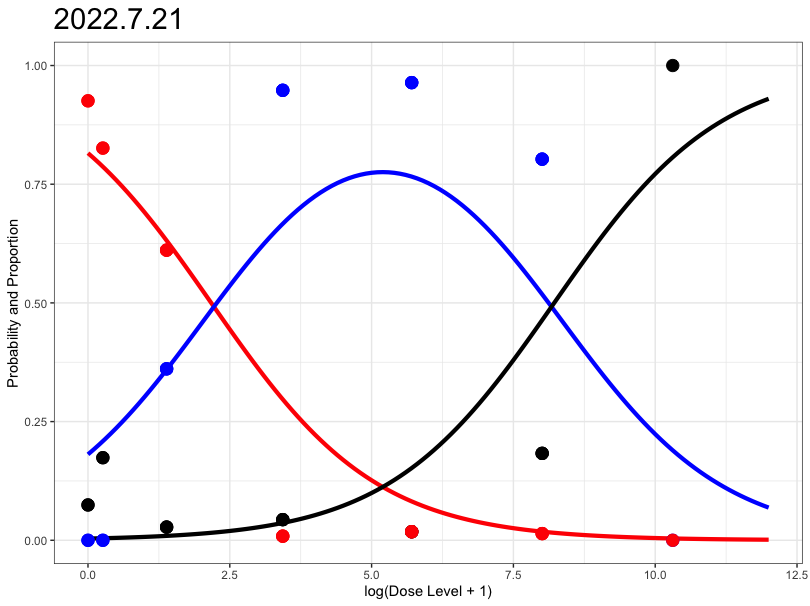}}
\subfloat[]{\includegraphics[width=0.32\textwidth]{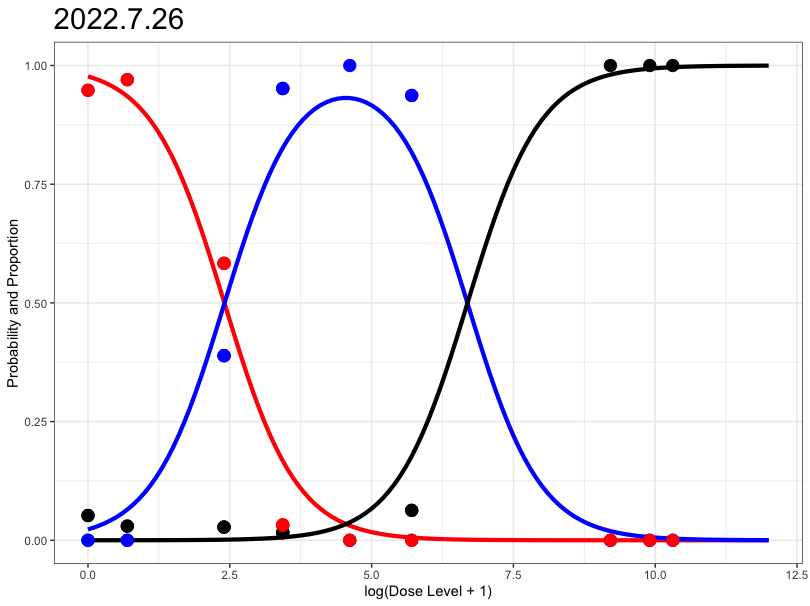}}  \\
 \subfloat[]{\includegraphics[width=0.32
 \textwidth]{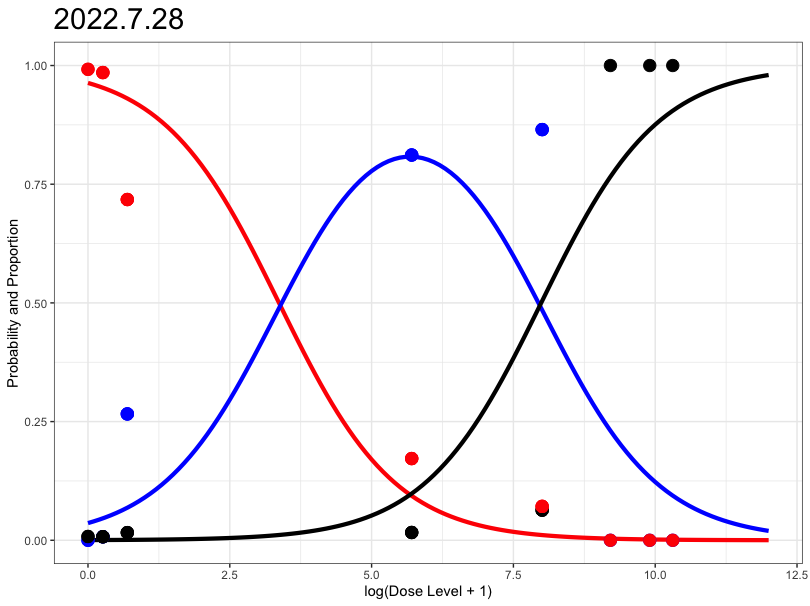}}
\subfloat[]{\includegraphics[width=0.32\textwidth]{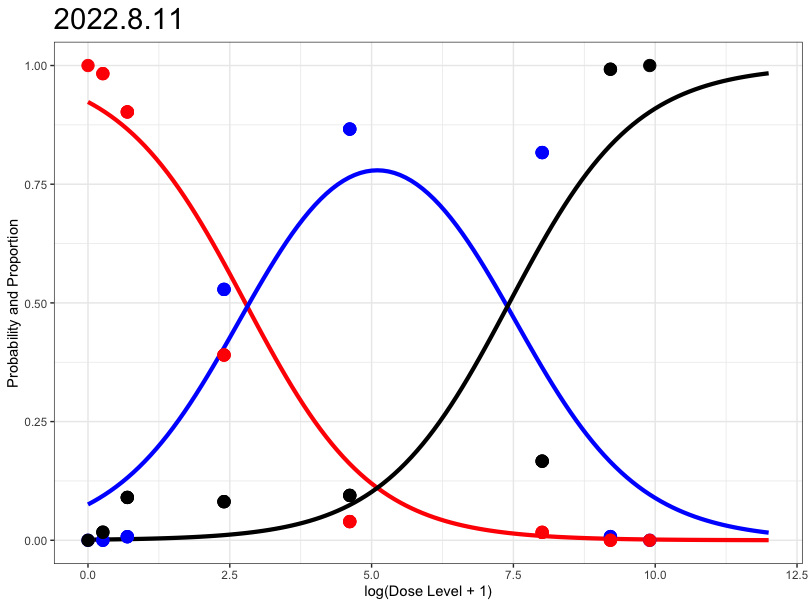}}  
 \subfloat[]{\includegraphics[width=0.32
 \textwidth]{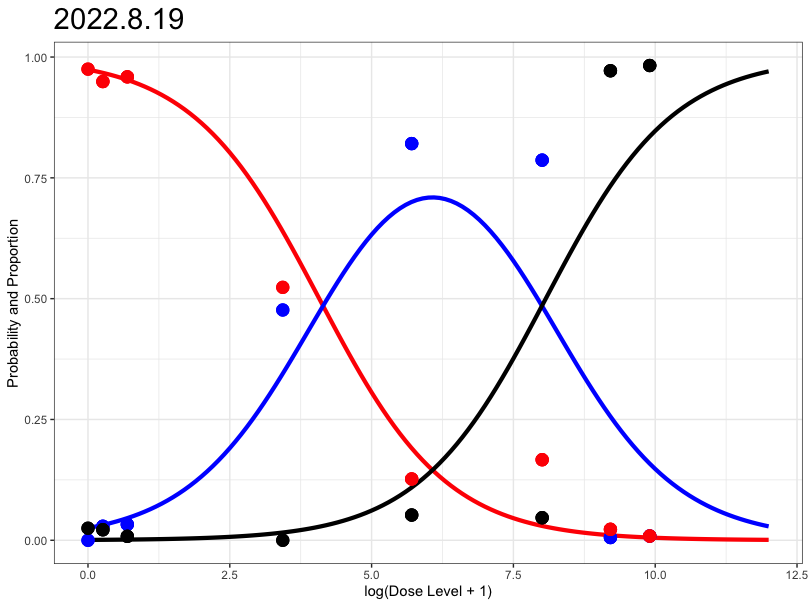}}
\caption{Daily fitted proportional odds models with logit link based on the first dataset. Red represents the observed and predicted proportion of normal embryos; blue represents the observed and predicted proportion of radial embryos; black represents the observed and predicted proportion of dead/delayed embryos.}\label{fig:otr_daily_fit}
\end{figure}

Each experiment records the number of embryos across four outcome categories: normal, radialization, 0 spicules, and dead/delayed. The experimental design points (dose levels) and corresponding weights varied significantly across the nine days, as summarized in Table~\ref{table:daily_eff}. 
\begin{table}[ht!]
\centering
\begin{tabular}{||r p{5cm} p{5cm} c c||} 
 \hline
Date & Design points & Design weights &	$\text{Eff}_D$ & $\text{Eff}_c$ \\ [1ex] 
 \hline\hline
  06/23 & [0, 1, 5, 10, 30, 100] &  [0.16, 0.17, 0.17, 0.19, 0.17, 0.16] & 0.448 & 0.640\\[1ex] 
  07/07 &  [0, 100, 300, 1000] &  [0.25, 0.25, 0.25, 0.25] & 0.755 & 0.199 \\ 
  07/14 &  [0, 1, 3, 10, 30, 100, 300, 3000, 10000] & [0.12, 0.12, 0.1, 0.11, 0.11, 0.11, 0.11, 0.10, 0.12] & 0.900 & 0.430\\
  07/19 & [0, 3, 10, 30, 100, 300, 10000, 30000] & [0.13, 0.12, 0.12, 0.14, 0.11, 0.12, 0.15, 0.11] & 0.869 & 0.425 \\
  07/21 & [0, 0.3, 3, 30, 300, 3000, 30000] &  [0.14, 0.16, 0.13, 0.13, 0.13, 0.17, 0.14] & 0.830 & 0.317 \\
  07/26 & [0, 1, 10, 30, 100, 300, 10000, 20000, 30000] & [0.11, 0.13, 0.10, 0.12, 0.11, 0.11, 0.11, 0.11, 0.11] & 0.843 & 0.356 \\ [1ex]
  07/28 & [0, 0.3, 1, 30, 3000, 10000, 20000, 30000] &  [0.13, 0.14, 0.12, 0.12, 0.13, 0.14, 0.11, 0.12] & 0.692 & 0.141 \\
  08/11 & [0, 0.3, 1, 10, 100, 3000, 10000, 20000] & [0.11, 0.12, 0.14, 0.13, 0.13, 0.12, 0.13, 0.12] & 0.814 & 0.305 \\
  08/19 & [0, 0.3, 1, 30, 300, 3000, 10000, 20000] & [0.11, 0.13, 0.11, 0.12, 0.12, 0.14, 0.16, 0.11] & 0.836 & 0.249 \\
 \hline
\end{tabular}
\caption{$D$- and $c$-efficiencies of designs used in the first dataset.}
\label{table:daily_eff}
\end{table}
We fitted multiple regression models to analyze the data, including ordinal regression (proportional odds), continuation ratio, and adjacent categories logit models. Model selection was guided by AIC and BIC values (Table~\ref{tab:tri_model_comp}).
\begin{table}[]
    \centering
    \begin{tabular}{l|cc}
    \hline\hline
    Model & AIC & BIC\\
    \hline
      Cumulative logit model   & 7212.557 & 7221.376\\
      Proportional odds model with Cauchit link & 7973.305 & 7994.327 \\
       Proportional odds model with logit link  & 6747.137 & 6768.160\\
       Adjacent categories logit model & 7720.505 & 7727.119\\
       Continuation-ratio logit model & 7656.651 & 7663.265\\
       \hline\hline
       
    \end{tabular}
    \caption{Comparison of trinomial models}
    \label{tab:tri_model_comp}
\end{table}
\textcolor{black}{Based on the AIC and BIC results, we choose to use proportional odds model with logit link for fitting the 9 day data separately. The fitted proportional odds model with logit link using the whole first dataset is given in Figure~\ref{fig:fitted_ordinal_together} while the fitted models using 9 day data separately is given in Figure~\ref{fig:otr_daily_fit} and estimated parameters are given in Table~\ref{tab:params}. Later we utilize the estimated parameters to construct the so-called robust designs (Section~\ref{sec:two_stage_robust}) \citep{collins2022model}.}

The proportional odds model with logit link was selected for its superior performance. Figure~\ref{fig:otr_daily_fit} illustrates the fitted models for each experimental day, highlighting variations in dose-response relationships due to biological variability.

\begin{figure}
    \centering
    \includegraphics[scale=0.2]{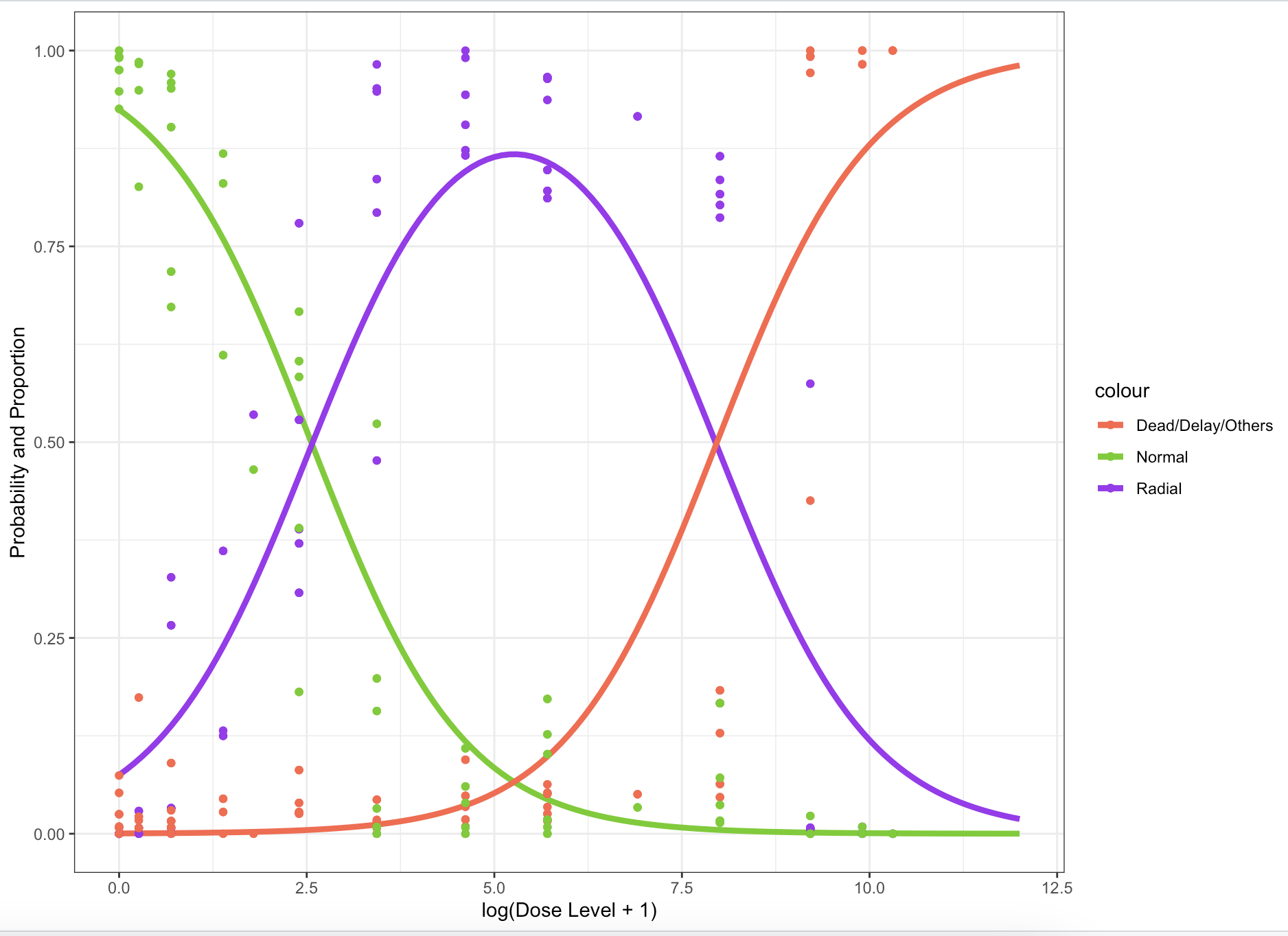}
    \caption{The fitted proportional odds model with logit link using the whole first dataset.}
    \label{fig:fitted_ordinal_together}
\end{figure}

Although the data is not independent across 9 days, the resulting dose-response curve can still provide invaluable information when we have different designs. Hence, the resulting second stage design stage is expected to be more robust against the design that we only use one set of nominal values (i.e., the one that we fit the 9-day data all altogether). Taking the first panel in Figure~\ref{fig:otr_daily_fit} as an example, the dose levels are low compared with others, reflecting potential genetic variants in sea urchins across different timelines (since the second stage design is performed at a different time compared with the first one).

    \begin{table}[h]
\centering
\begin{tabular}{c|ccc}
\hline\hline
Date & $\beta_1$ & $\beta_2$ & $\alpha$ \\ \hline
06/23 & 2.328 & 9.845 & -1.562 \\ 
07/07 & 2.077 & 10.686 & -1.303 \\
07/14 & 2.157 & 9.342 & -1.019 \\
07/19 & 2.516 & 9.127 & -1.086 \\
07/21 & 2.186 & 8.029 & -0.960 \\
07/26 & 2.380 & 8.359 & -1.040 \\
07/28 & 2.442 & 8.331 & -1.037 \\
08/11 & 2.449 & 8.121 & -1.021 \\
08/19 & 2.506 & 7.800 & -0.979 \\ \hline\hline
\end{tabular}
\caption{9 sets of nominal values based on the first dataset.}
\label{tab:params}
\end{table}

\begin{figure}
    \centering
    \includegraphics[scale=0.2]{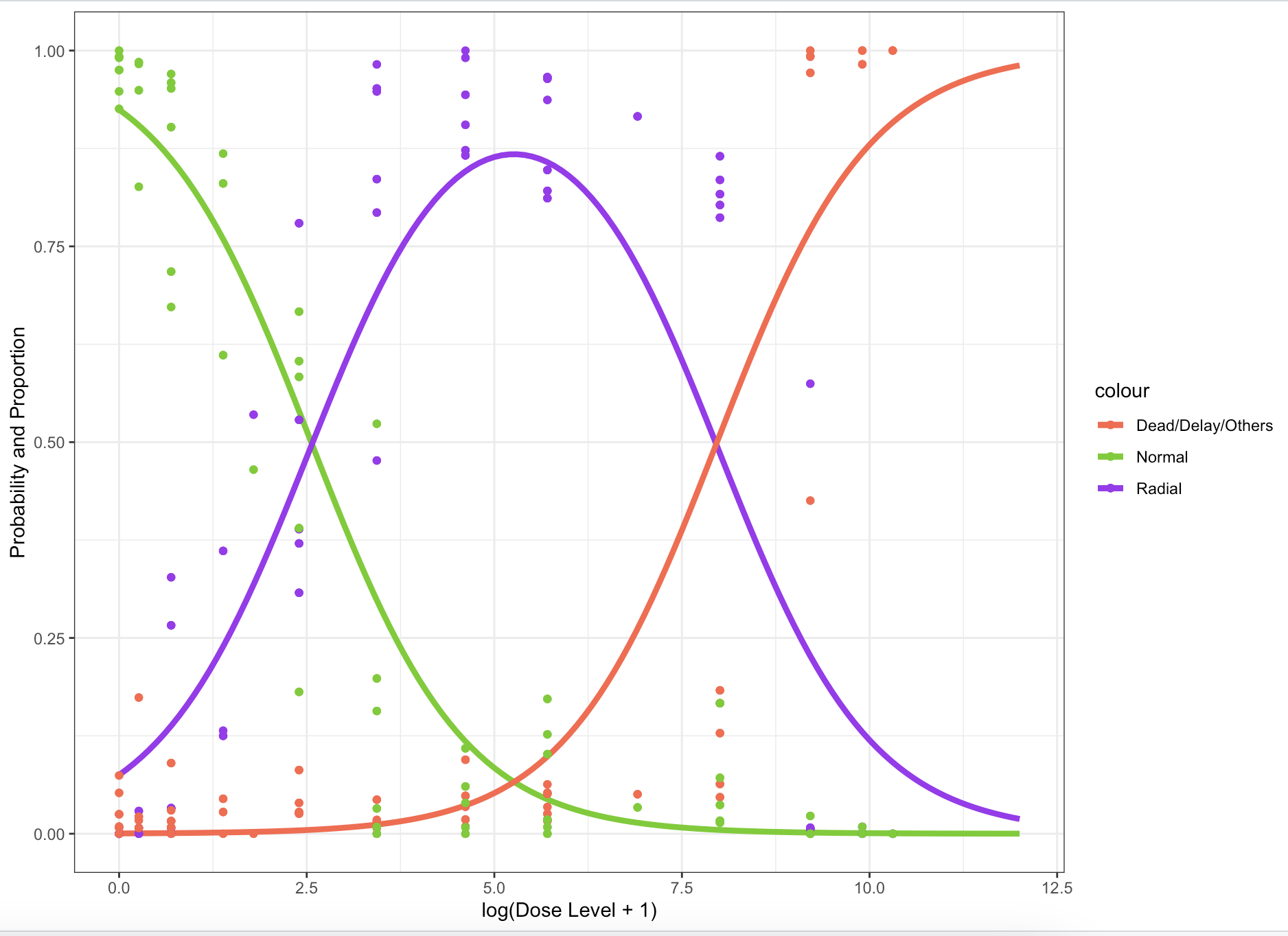}
    \caption{The fitted proportional odds model with logit link using the whole first dataset.}
    \label{fig:fitted_ordinal_together}
\end{figure}

\subsubsection{The Second Dataset: Locally Optimal Design}\label{sec:locally_optimal}
We construct a locally optimal design to compare it with our proposed two-stage robust design. To construct locally optimal design, we need to first fit the whole dataset and then use the nominal values (i.e., fitted parameter estimates) to compute an optimal design numerically. To estimate the parameters $\theta=(\beta_1,\beta_2,\alpha)$ as accurate as possible, we consider minimizing the volume of the confidence set:
$$S=\left\{\theta:(\widehat{\theta}-\theta)^TM^{-1}(\widehat{\theta}-\theta)\le \chi^2_{df,0.95}\right\}$$
where $\widehat{\theta}$ is the estimator of $\theta$, $M$ is the asymptotic variance of $\widehat{\theta}$ and $\xi^2_{df,0.95}$ is the $95\%$-quantile of Chi-square distribution with degrees of freedom $df$. It can be shown that the volume of $S$ is proportional to the determinant of $M^{-1}$, which is the $D$-optimality in literature.

In addition, RD50 is another important quantity of interest (personal communication with Dr. Collins). Hence, minimizing the asymptotic variance of an estimator of RD50 is another goal of the design. The asymptotic variance of RD50 has a neat form and we call it the $c$-optimality. Then a dual-optimality criteria can be constructed as a convex combination of $D$- and $c$-optimality. 

\begin{figure}[!htbp]
\centering
{\includegraphics[width=0.55\textwidth]{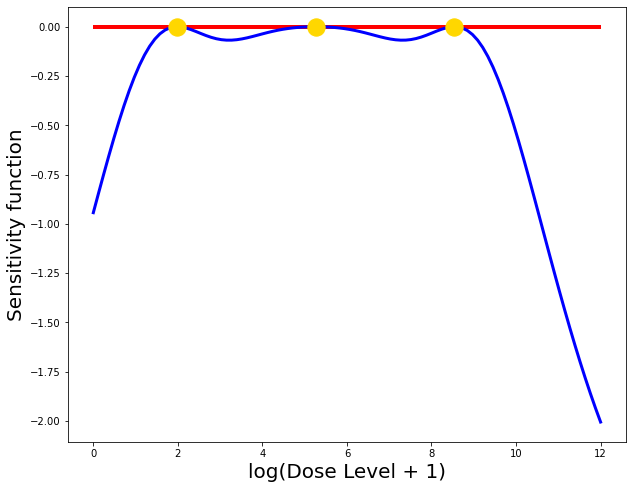}}
\caption{Sensitivity function of the dual-optimal design. \textcolor{black}{The x-axis represents the log-transformed dose levels, while the y-axis shows the sensitivity function values. The blue curve represents the sensitivity function across these dose levels, while the red line at the top of the plot highlights the zero-sensitivity baseline.}}\label{fig:dual_sens}
\end{figure}

The dual-optimal design is then defined as
\begin{align}\label{eq:dual_optim}
    \xi_{dual}&=\arg\min_{\xi\in\Xi}\Phi_{dual}(M)=\arg\min_{\xi\in\Xi} \left(\lambda\frac{\Phi_D(M)}{3}-(1-\lambda)\log\Phi_c(M)\right)
\end{align}
where $\Phi_D(M)=\log\det M$ and $\Phi_c(M)=\left(\nabla_{\theta}x|_{\theta=\widehat{\theta}}\right)^TM(\xi)^{-1}\left(\nabla_{\theta}x|_{\theta=\widehat{\theta}}\right)$ and $\lambda\in[0,1]$ is a weight to trade-off between $D$- and $c$-optimality. The optimization problem~\ref{eq:dual_optim} is constrained because the design space is $[0,+\infty)$ and the design weights are all within $[0,1]$ with summation 1. Hence, we apply Particle Swarm Optimization (PSO) \citep{miranda2018pyswarms} to solve it with the special choice $\lambda=0.5$, indicating an equal importance of both criteria. Using the estimated parameters of the first dataset in table~\ref{table:est_otr}, the resulting locally dual-optimal design with equal weights and additional zero dose level is given in the table~\ref{table:dual_optim} which forms the basis of the second dataset. The sensitivity function is shown in figure~\ref{fig:dual_sens}. Note that the table shows the raw scale of dose level while for implementation, we use $\log(\text{Dose level} + 1)$ so that the design space is the same as the original space $[0,+\infty)$ but the numerical issues can be lightened.

\begin{table}[ht!]
\centering
\begin{tabular}{||r p{5cm} p{5cm} c c||} 
 \hline
Date & Design points & Design weights &	$\text{Eff}_D$ & $\text{Eff}_c$ \\ [1ex] 
 \hline\hline
 $\xi_D$ & [5.77, 161.4, 4391.52] &  [0.33, 0.34, 0.33] & 1.000 & 0.372\\
 $\xi_c$ &  [12.01] & [1.00] & Singular & 1.000\\
 $\xi_{dual}$ & [9.08, 59.7, 4290] & [0.65, 0.175, 0.175] & 0.798 & 0.748 \\
  \hline\hline
  12/03 & [0, 9.05, 59.7, 4290] & [0.169, 0.521, 0.169, 0.169] & 0.767 & 0.688 \\
  12/09 & [0, 9.08, 59.7, 4290] & [0.169, 0.521, 0.169, 0.169] & 0.767 & 0.688 \\
  12/10 & [0, 9.08, 59.7, 4290] & [0.169, 0.521, 0.169, 0.169] & 0.767 & 0.688 \\
 \hline
\end{tabular}
\caption{$D$- and $c$-efficiencies of the December data.}
\label{table:dual_optim}
\end{table}

The estimated parameters and their standard errors based on the second dataset (also called the December data) is given in the right column of Table~\ref{table:est_otr} and the fitted dose-response curve is given in Figure~\ref{fig:prop_odds_dec}. It is obvious that the curve does not fit the data well and the corresponding standard errors are not small enough as expected (though we are using dual-optimality here). Hence, it is not enough to just use optimal design alone, and we are expecting that the two-stage design may perform better. \textcolor{black}{To investigate the gap between the reality and theory in depth, we conjecture that there could a genetic shift in the dataset (see also Figure~\ref{fig:otr_daily_fit}), meaning that the tolerance of sea urchin changes in the genetic composition of populations over time, often in response to environmental pressures such as temperature, pH, or salinity. In sea urchins, like many marine organisms, these shifts can influence key developmental processes in embryos and can impact the resilience of populations to climate change and ocean acidification \citep{jorde2004genetic}.}

\begin{table}[h!]
\centering
\begin{tabular}{||c c c||} 
 \hline
 & The first dataset (June-August) &	The second dataset (December) \\ [0.5ex] 
 \hline\hline
  Total observations & 8163 & 2070 \\
 $\beta_1$ & 2.506 (0.055) & 0.593 (0.081)  \\ [1ex] 
 $\beta_2$ & 7.800 (0.134) & 6.106 (0.231)  \\ [1ex] 
 $\alpha$ & 0.979 (0.016) & 0.719 (0.030)  \\ [1ex] 
 $RD50$ & 2.580 (0.089) & 0.780 (0.149)  \\ [1ex] 
 \hline
\end{tabular}
\caption{Ordinal regression models based on two datasets.}
\label{table:est_otr}
\end{table}

\subsubsection{The Third and Fourth Datasets: Two-stage Robust Optimal Design}\label{sec:two_stage_robust}
The name "robust" comes from the fact that we use different sets of nominal values to construct the optimal design \citep{zang2005review, collins2022model}.The optimal design is constructed as follows:
$$\xi_{opt}=\arg\max_{\xi}\sum_{i=1}^K\Phi(\xi,\widehat{\theta}_i)$$
where each $\Phi(\xi,\theta_i)$ is a criterion function that uses the nominal value $\theta_i$. In the case of D-optimality, $\Phi(\xi,\theta_i)$ is the log determinant of Fisher information matrix while in the case of dual-optimality for estimating parameters and oral radialization 50\%, $\Phi(\xi,\theta_i)$ is the convex combination of log D-efficiency and log c-efficiency \citep{hyun2015multiple}.

We construct the two-stage robust optimal designs based on the first dataset. To achieve this, we need to determine how many additional doses we want add to the first dataset. In our real application, we set this as the same number of total observations as the first dataset ($\alpha=0.5$). Further, we determine to put a fixed proportion to zero dose level (the control group) and 10,000 dose level (the endpoint to make sure all sea urchins are dead). The resulting two-stage robust D-optimal design is given in Table~\ref{table:2stage_augmented_robust_d_optim} while the two-stage robust dual-optimal design (with equal weights) is given in Table~\ref{table:2stage_augmented_robust_dual_optim}.

\begin{table}[h!]
\centering
\begin{tabular}{||c c c c c c c||} 
 \hline
Design points & 0 &	14 &55& 683 & 4808&10000  \\ [1ex] 
 \hline
  Design weights &  0.225  &	0.145	& 0.112	&0.151 &	0.142	&0.225 \\[1ex] 
 \hline
\end{tabular}
\caption{Two-stage robust D-optimal design based on the first dataset.}
\label{table:2stage_augmented_robust_d_optim}
\end{table}

\begin{table}[h!]
\centering
\begin{tabular}{||c c c c c c c||} 
 \hline
Design points & 0 &	5 & 25 & 989 & 3727 & 10000  \\ [1ex] 
 \hline
  Design weights & 0.215 & 0.200 & 0.305 & 0.033 & 0.032 & 0.215 \\[1ex] 
 \hline
\end{tabular}
\caption{Two-stage robust dual-optimal design with equal weights based on the first dataset.}
\label{table:2stage_augmented_robust_dual_optim}
\end{table}

\begin{figure}
    \centering
    \begin{subfigure}[b]{0.3\textwidth}
    \centering
    \includegraphics[width=\textwidth]{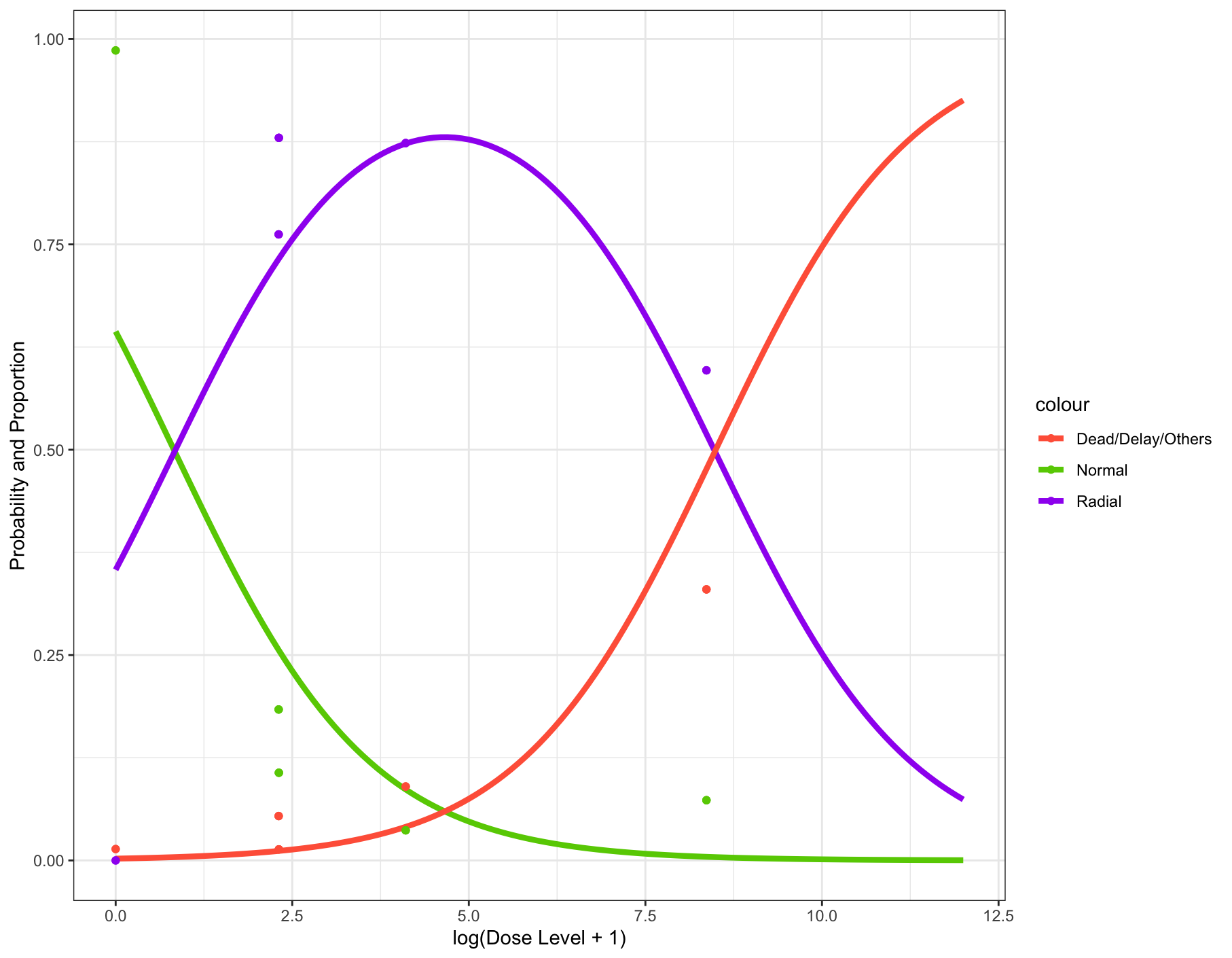}
    \caption{Fitted proportional odds model based on the December data (the second dataset).}
    \label{fig:prop_odds_dec}
\end{subfigure}
\hfill
    \begin{subfigure}[b]{0.3\textwidth} 
        \centering
        \includegraphics[width=\textwidth]{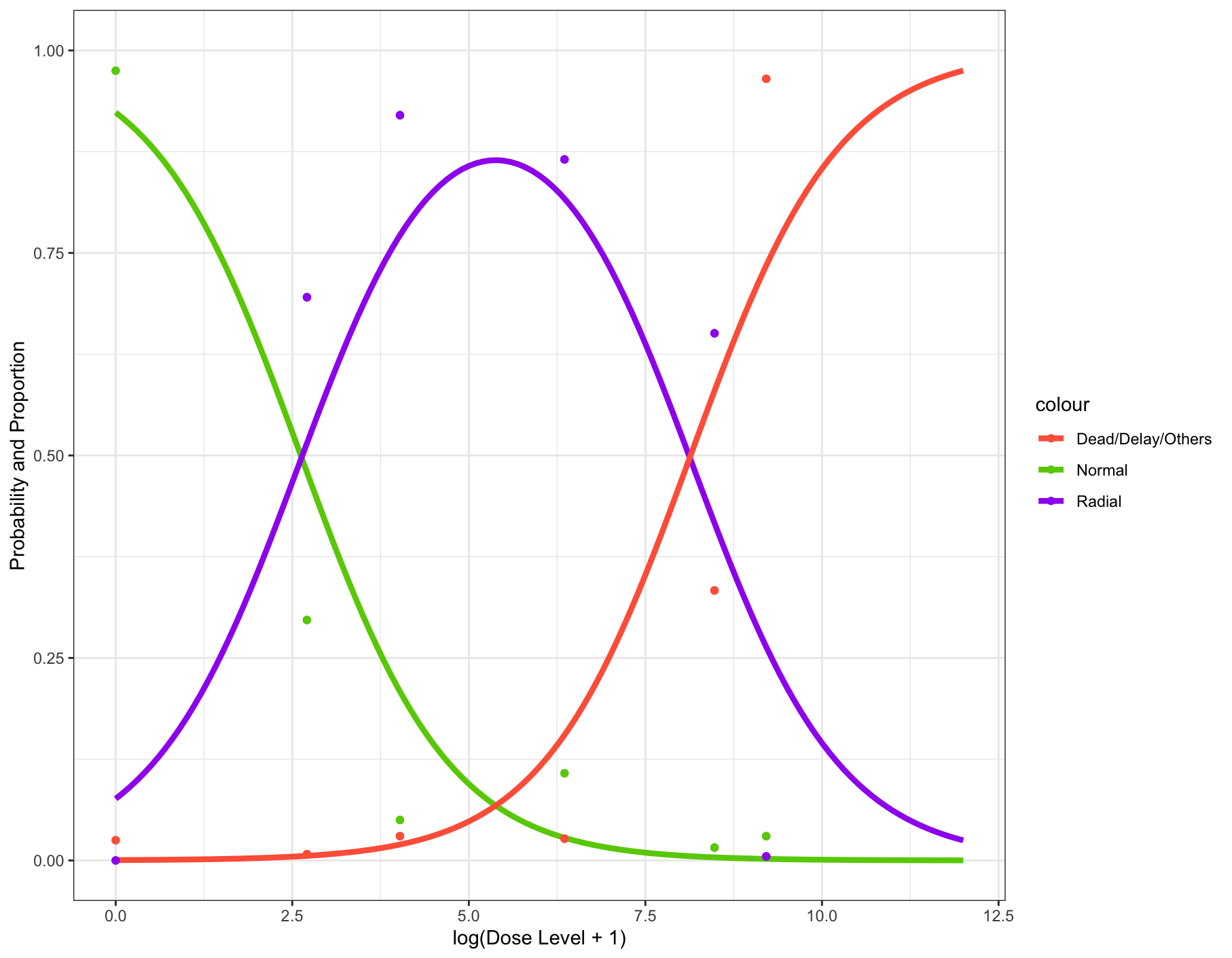}
        \caption{Fitted dose-response curve based on the two-stage robust D-optimal design (the third dataset).}
        \label{fig:sub1}
    \end{subfigure}
    \hfill
    \begin{subfigure}[b]{0.3\textwidth} 
        \centering
        \includegraphics[width=\textwidth]{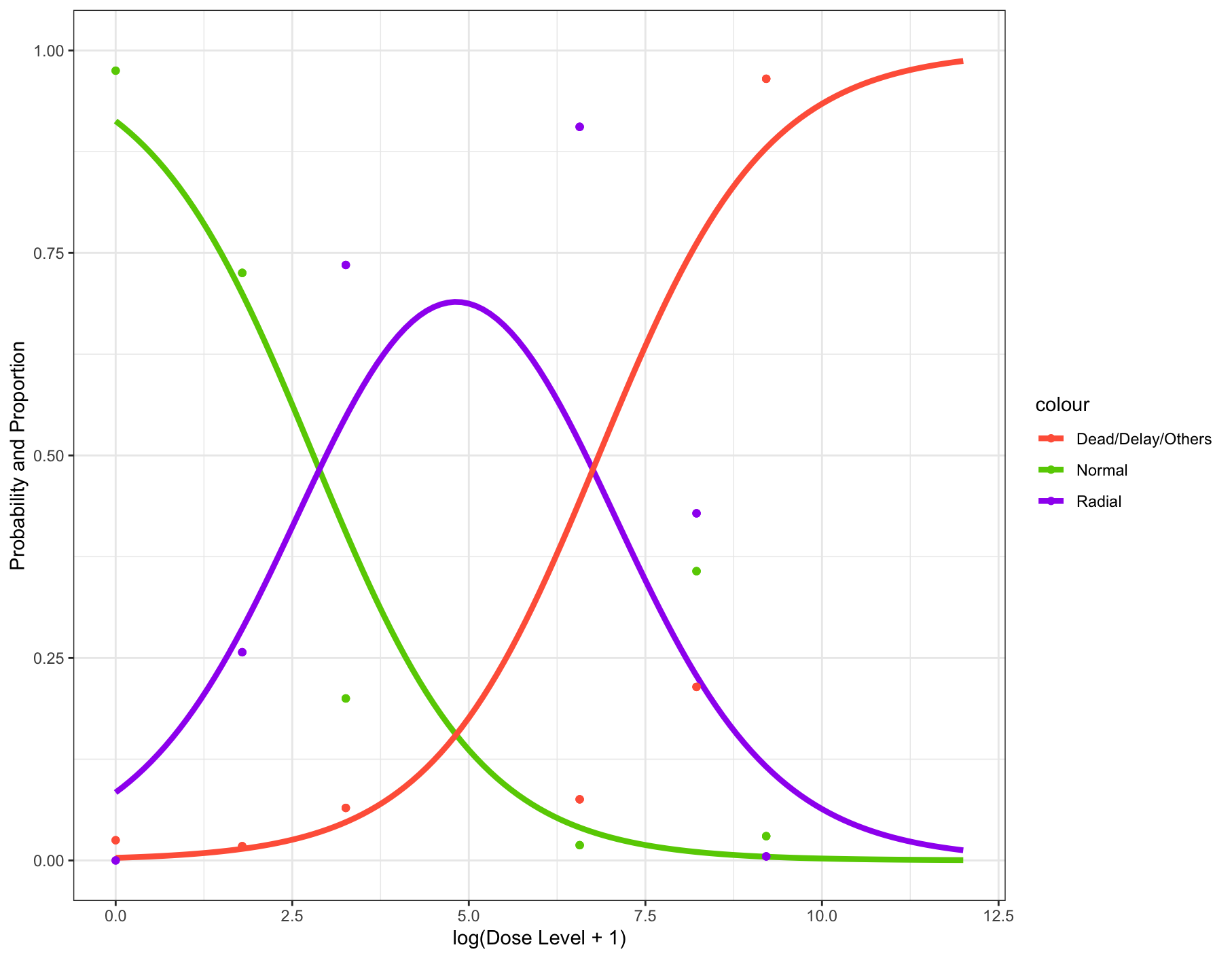}
        \caption{Fitted dose-response curve based on the two-stage robust Dual-optimal design (the fourth dataset).}
        \label{fig:sub2}
    \end{subfigure}
    \caption{ \textcolor{black}{Dose-response curves fitted using various datasets and design strategies, illustrating the probability (or proportion) of different outcomes across dose levels on a logarithmic scale. The curves represent three outcome categories: "Dead/Delayed/Others" (red), "Normal" (green), and "Radial" (purple). Plot (a) shows the dose-response curve using a proportional odds model based on the December data, representing a conventional approach. Plot (b) uses the two-stage robust D-optimal design, adjusting the dose-response curve to improve model robustness. Plot (c) applies the two-stage robust dual-optimal design, further refining the dose-response prediction, particularly for the "Radial" and "Dead/Delayed/Others" categories, and demonstrating the enhanced adaptability of the dual-optimal approach. These variations highlight the effects of different optimization criteria on dose-response predictions across varying outcome probabilities.}}
    \label{fig:robust_design}
\end{figure}

Based the above suggested designs, we collect additional toxicological data and we have the following four datasets:
\begin{itemize}
    \item The first dataset from June-August;
    \item The second dataset in December which is based on the first dataset and is dual-optimal (not robust);
    \item The third dataset which is based on the two-stage robust D-optimal design;
    \item The fourth dataset which is based on the two-stage robust dual-optimal design.
\end{itemize}
The results of fitted ordinal regression models for the third and fourth datasets are shown in Table~\ref{table:twostage_design} and  Figure~\ref{fig:robust_design}.  \textcolor{black}{Table~\ref{table:twostage_design} provides fitted parameters from ordinal regression models with a logit link function using two robust design strategies: the two-stage robust D-optimal and the two-stage robust dual-optimal designs. Here, we note that the total number of observations is slightly higher for the dual-optimal design (927 compared to 889 for the D-optimal design), as the actual number of observations may fluctuate a little in contrast to theoretical calculations. Parameter estimates, such as \(\beta_1\), \(\beta_2\), and \(\alpha\), as well as the RD50, are provided along with standard errors. The smaller standard errors in the dual-optimal design indicate potentially greater precision for this design. Interestingly, the standard errors of the dual-optimal design are generally lower than the D-optimal designs, which contradicts with the intuition and the original intention of both designs. We further explore this issue in Table~\ref{tab:compare_four_datasets}. In addition, the RD50 estimate is slightly higher under the dual-optimal design (2.955 compared to 2.650), which could imply differences in dose-response sensitivity captured by each design.}

\textcolor{black}{
Table~\ref{tab:compare_four_datasets} provides a detailed look at the $D-$ and $c-$ optimality values and their efficiencies under different design settings. The efficiencies are computed in a relative manner, i.e., it is the actual efficiency divided by the largest efficiency among 4 designs and sample sizes (total number of observations) are scaled so results from different designs are comparable. Here $\Theta_i,i=1,2,3,4$ correspond to the parameters that we estimated from four different designs. The Conventional design refers to the original dataset designed by Dr. Michael Collins; the Dual, Robust-D and Robust Dual designs refer to the locally dual-optimal, robust-D and robust-dual designs based on $\Theta_1$ and daily fitted models (see Figure~\ref{tab:params}) respectively.  Interestingly and sarcastically, the Robust D design never performs the best in terms of D-optimality among all scenarios ($\Theta$'s). One most plausible explanation is still due to the genetic shift of the dataset, making it awkward for the new dataset it has collected. Apart from that, all designs perform as expected and in some scenarios the Dual design performs better than the Robust Dual design in terms of both $D$- and $c$-optimality, suggesting that there is always a gap between theory and practice.
}

\begin{table}[h!]
\centering
\begin{tabular}{||c c c||} 
 \hline
 & Two-stage robust D-optimal design &	Two-stage robust Dual-optimal design \\ [0.5ex] 
 \hline\hline
  Total observations & 889 & 927 \\
 $\beta_1$ & 2.694 (0.181) & 2.065 (0.127)  \\ [1ex] 
 $\beta_2$ & 7.449 (0.430) & 5.313 (0.298)  \\ [1ex] 
 $\alpha$ & 0.906 (0.043) & 0.787 (0.040)  \\ [1ex] 
 $RD50$ & 2.650 (0.154) & 2.955 (0.123)  \\ [1ex] 
 \hline
\end{tabular}
\caption{Fitted ordinal regression models with logit link based on two-stage robust designs.}
\label{table:twostage_design}
\end{table}

\begin{table}[h!]
\centering
\begin{tabular}{l|p{2.5cm}|rrrr}
\hline\hline
Estimates &Datasets& 1 (Conventional) & 2 (Dual) & 3 (Robust D) & 4 (Robust Dual) \\
\hline
$\Theta_1$ & D-optimality & 5.918&	6.293&	6.118	&6.417 \\
 & D-efficiency & 1.000&	0.882&	0.935&	0.847\\
 & c-optimality & 11.990&	6.305&	17.103&	10.191\\
 & c-efficiency &0.526&	1.000&	0.369&	0.619 \\
\hline
$\Theta_2$ & D-optimality & 4.997&	5.556&	5.039	&5.163 \\
 & D-efficiency & 1.000&	0.830&	0.986&	0.946 \\
 & c-optimality & 17.961	&17.160	&18.202&	13.753\\
 & c-efficiency & 0.766	&0.801&	0.756&	1.000\\
\hline
$\Theta_3$ & D-optimality & 5.783&	6.159&	6.025	&6.386 \\
 & D-efficiency & 1.000	&0.882&	0.923&	0.818 \\
 & c-optimality & 12.582	&7.048	&18.508&	12.106 \\
 & c-efficiency & 0.560&	1.000&	0.381&	0.582 \\
\hline
$\Theta_4$ & D-optimality & 4.891	&5.063&	5.202&	5.353\\
 & D-efficiency & 1.000&	0.944	&0.902&	0.857 \\
 & c-optimality & 14.891&	8.943&	19.921&	12.787\\
 & c-efficiency &0.601	&1.000&	0.449&	0.699 \\
\hline\hline
\end{tabular}
\caption{Comparison among four datasets. $\Theta_1,\Theta_2,\Theta_3,\Theta_4$ stand for estimated parameters $(\beta_1,\beta_2,\alpha)$ from the four different datasets.}
\label{tab:compare_four_datasets}
\end{table}

\subsubsection{Equivalence Theorems}\label{sec:equi_new}
In this subsubsection, we derive the equivalence theorems for the two-stage robust dual-optimal design with an additional 0 dose level under the ordinal regression setting. 

We start with a more general case which is the multivariate logistic regression. The multivariate logistic regression is an extension of the usual logistic regression for categorical outcomes with or without intrinsic ordering (personal communication with Dr. Weng Kee Wong). Suppose $\m{y}\sim\mathcal{M}(n,\pi)$ and we assume
\begin{align*}
    \eta&=X\beta,\    \m{\pi}=\mu(\theta)\\
    \eta&=g(\mb{\pi})=\psi^{-1}(\mb{\pi})=C^T\log(L\mb{\pi})\\
   \theta&=r(\eta)=\log\mb{\pi}=\log g^{-1}(\eta)
\end{align*}
where $X$ is the design matrix and $C$ and $L$ are matrices of contrasts and marginal indicators \citep{glonek1995multivariate}. Different choices of $C$ and $L$ lead to different models such as proportional odds model, adjacent categories logit model and continuation ratio logit model. Further, the nominal logistic regression can also be written in this form \citep{glonek1995multivariate}. Most importantly, by introducing $L$ we ensure the mapping $\mb{\pi}\rightarrow\eta$ is of full rank so that we can write $\theta=\log\mb{\pi}$. Thus, we implicitly put the linear constraint $\m{1}^T\exp({\theta})=1$ so that in this case $A(\theta)=0$.

For convenience, we note:
\begin{itemize}
    \item $y$, $\mb{\pi}$, $\theta$ and $\eta$ are all $k\times 1$ vectors;
    \item $\beta$ is a $p\times 1$ vector;
    \item $X$ is a $k\times p$ matrix of full row rank;
    \item $L$ and $C$ are $l\times k$ matrices both of full column rank.
\end{itemize}

The Fisher information of $\beta$ is given by
    \begin{align}
        \m{M}(\beta)&=X^T\left(C^TD^{-1}L\right)^{-T} \m{V}^{-1} \left(C^TD^{-1}L\right)^{-1} X\\ 
        &=\widetilde{s}(X)\widetilde{s}(X)^T\nonumber
    \end{align}
    where $D=\text{Diag}(L\mb{\pi})$ and
    \begin{align*}
        \widetilde{s}(X)&=X^T\left(C^TD^{-1}L\right)^{-1} \m{V}^{-1/2},\\
        \m{V}^{-1/2}&=\text{Diag}(1/\sqrt{\mb{\pi}}).
    \end{align*}
    The $\widetilde{s}(X)$ representation is particularly useful we are deriving equivalence theorems and sensitivity functions in optimal design theory. More properties of multivariate logistic regression and its extensions are given in \cite{glonek1995multivariate, lang1996maximum, bu2020d}. 

\begin{example}[Ordinal Trinomial Regression \citep{zocchi1999optimum}]\label{example:OTR} Following the previous example, the $3$-dimensional response vector $y=(y_1,y_2,y_3)^T\sim\mathcal{M}(1,\m{\pi})$ has an intrinsic ordering, i.e., $y_1$, $y_2$ and $y_3$ correspond to normal, radialization and dead, respectively. We model $y$ using a proportional odds model with common slope:
\begin{align*}
    \eta_1&=\log\left(\frac{\pi_1}{\pi_2+\pi_3}\right)=\beta_1+\alpha x\\
    \eta_2&=\log\left(\frac{\pi_1+\pi_2}{\pi_3}\right)=\beta_2+\alpha x\\
    \eta_3&=\log(\pi_1+\pi_2+\pi_3)=0
\end{align*}
    In this case, the parameter $\theta$ and the design matrix $X$ are 
    \begin{align*}
    \theta&=\left(\begin{matrix}
        \beta_1&\beta_2&\alpha& 0
    \end{matrix}\right)^T,
        &X=\left(\begin{matrix}
            1& 0& x& 0\\
            0& 1& x& 0\\
            0& 0& 0& 1
        \end{matrix}\right)
    \end{align*}
    and the choices of $L$ and $C^T$ are 
    \begin{align*}
            L&=\left(\begin{matrix}
        1&0&0\\
        1&1&0\\
        0&1&1\\
        0&0&1\\
        1&1&1
    \end{matrix}\right),
    &C^T=\left(\begin{matrix}
        1&0&-1&0&0\\
        0&1&0&-1&0\\
        0&0&0&0&1
    \end{matrix}\right).
    \end{align*}
\end{example}

In a usual sequential design setting, we add a point $x$ to the existing design $\xi$ so that the corresponding design efficiency (say, $D$-efficiency) is improved. However, as we mentioned in Section~\ref{subsec:augmented}, this is not always what toxicologists want in practice. In practice, we always want a "control group" under any experimental design. In other words, there should always a $0$ dose level group. In this case, the design becomes
$$\xi=\left(\begin{matrix}
0&x_1&\cdots&x_n\\
\alpha&(1-\alpha) p_1&\cdots&(1-\alpha)p_n
\end{matrix}\right)$$
where $p_i,i=1,\cdots,n$ are nonnegative design weights and $\alpha\in[0,1]$ is another weight assigned to the control group $x=0$. In the following, we derive the expression of Fisher information when we add a new design point to the original design.

\begin{example}[Adding a new design point]\label{example:add_a_point} In example~\ref{example:OTR}, the Fisher information matrix is of form
$$M(\beta)=\widetilde{s}(X)\widetilde{s}(X)^T$$
where $\widetilde{s}(X)$ is a $p\times k$ matrix. Let $M_1=\sum_{i=1}^np_i\widetilde{s}(X_i)\widetilde{s}(X_i)^T$ be the information matrix associated with $\xi_1$ and $p_i$ are nonnegative design weights. Also let $M_0=\widetilde{s}(X_0)\widetilde{s}(X_0)$ be the information matrix at point $0$. If we want to add $0$ to the design $\xi_1$ with weight $\alpha\in(0,1)$, then by Sherman-Woodbury-Morrison, the corresponding $D$-optimality criteria becomes 
\begin{align}
    \det\left((1-\alpha)M_1+\alpha M_0\right)&=(1-\alpha)^p\det(M_1)\det\left(I+\frac{\alpha}{1-\alpha}\widetilde{s}(X_0)^TM_1^{-1}\widetilde{s}(X_0)\right).
\end{align}
Further, the $c$-optimality for estimating $g(\theta)$ becomes
\begin{align}
    &\nabla g^T\left((1-\alpha)M_1+\alpha M_0\right)^{-1}\nabla g =\nonumber\\&\frac{\nabla g^TM_1^{-1}\nabla g}{1-\alpha}-\frac{\alpha}{(1-\alpha)^2}\nabla g^TM_1^{-1}\widetilde{s}(X_0)\left(I+\frac{\alpha}{1-\alpha}\widetilde{s}(X_0)^TM_1^{-1}\widetilde{s}(X_0)\right)^{-1}\widetilde{s}(X_0)^TM_1^{-1}\nabla g
\end{align}
where $\nabla g$ is the gradient of $g(\theta)$ at $\theta$.
\end{example}
Following example~\ref{example:OTR}, the Fisher information matrix of $\beta$ associated with $\xi$ is $M(\xi)=\alpha s_0s_0^T+(1-\alpha) M_1$ where $s_0=\widetilde{s}(X)|_{x=0}$ and $M_1$ is the Fisher information matrix associated with the design with point $0$ removed. Let us consider $D$-optimality, i.e., $\Phi_D(M)=\log\det M$, then by example~\ref{example:add_a_point}, we have
\begin{align}
    \Phi_D(M)&=\log\det(M_1)+p\log(1-\alpha)+\log\det\left(I+\frac{\alpha}{1-\alpha}s_0^TM_1^{-1}s_0\right).
\end{align}
Suppose $\alpha$ is fixed so that $M_1$ is the only free variable in the expression and we write $\Phi_D(M_1)$ instead of $\Phi_D(M)$. By standard matrix calculus, the Fréchet derivative of $\Phi_D(M)$ at $M_1$ in the direction of $M_2$ is
\begin{align}\label{eq:equi_D}
    F_{\Phi_D}(M_1, M_2)&=(1-\alpha)\text{Tr}\left(M^{-1}(M_2-M_1)\right)\\
    &=\text{Tr}\left(M_1^{-1}M_2-I\right)-\nonumber\\&\ \ \ \ \text{Tr}\left(\left(\frac{1-\alpha}{\alpha}I+s_0^TM_1^{-1}s_0\right)^{-1}\left(s_0^T(M_1^{-1}M_2M_1^{-1}-M_1^{-1})s_0\right)\right).
\end{align}
Now the general equivalence theorem applies using the fact that $\Phi_D(M_1)$ is concave in $M_1$. For c-optimality, we need to use the implicit function theorem \citep{rudin1976principles} to compute the asymptotic variance of, say, radialization 50\% estimate.

\begin{example}[Radialization 50\%]\label{example:radial}
\textcolor{black}{In this example, we are interested in finding a specific dose level of a substance that leads to a 50\% reaction rate (referred to as RD50), meaning it causes a measurable response in 50\% of the population. Using a trinomial regression model, which is a statistical method for analyzing outcomes that can take on three possible values, we aim to calculate this dose level. However, there isn’t a simple formula to solve for this dose directly based on the model parameters. Instead, we use mathematical techniques to approximate the effect of small changes in the model parameters on our estimate of the RD50 dose. This process allows us to calculate an "asymptotic variance," or a measure of how accurate and stable our estimated RD50 is likely to be with large sample sizes. Additionally, we look at a related measure, LD50, which indicates a 50\% lethal dose, and compute the variability for a ratio of these two metrics. This approach is valuable because it helps us assess the reliability of the dose estimates used in toxicological studies.}

Suppose we have the trinomial regression model (Example~\ref{example:OTR}) with parameters $\theta^T=(\alpha,\beta_1,\beta_2, b)$ and dose level $x$ (with the implicit constraint $b$=0). Suppose the interest is in estimating radialization $50\%$ (RD50), i.e., the dose level $x$ such that $\pi_2=0.5$. An equation for solving $x$ is
$$\frac{1}{1+\exp(-\eta_2)}-\frac{1}{1+\exp(-\eta_1)}-0.5=0.$$
There is no closed form solution of $x$ in terms of $\theta$. However, it is more important for us to derive the partial derivatives $\partial x/\partial\theta^T$ so that $\Delta$-method can be applied to approximate the asymptotic variance of estimated radialization $50\%$. Denote the left hand side of the equation as $f(x,\theta)$, then by the implicit function theorem, we have
\begin{align*}
A&=f'(x,\theta)\\
&=\left(\begin{matrix}
 \frac{\partial f}{\partial x} &
  \frac{\partial f}{\partial \alpha} &
   \frac{\partial f}{\partial \beta_1} &
    \frac{\partial f}{\partial \beta_2} &
     \frac{\partial f}{\partial b}
\end{matrix}\right)    \\
&=\left(\begin{matrix}
 \frac{\alpha\exp(-\eta_2)}{(1+\exp(-\eta_2))^2}- \frac{\alpha\exp(-\eta_1)}{(1+\exp(-\eta_1))^2}\\
 \frac{x\exp(-\eta_2)}{(1+\exp(-\eta_2))^2}- \frac{x\exp(-\eta_1)}{(1+\exp(-\eta_1))^2}\\
 -\frac{\exp(-\eta_1)}{(1+\exp(-\eta_1))^2}\\
 \frac{\exp(-\eta_2)}{(1+\exp(-\eta_2))^2}\\
 0
\end{matrix}\right)^T
\end{align*}
so that the desired $\nabla_\theta x$ can be computed numerically by plug-in estimates of $\theta$, i.e.,
$$\widehat{\nabla_\theta x}^T=\left(\frac{\partial f/\partial\alpha}{\partial f/\partial x},\frac{\partial f/\partial\beta_1}{\partial f/\partial x}, \frac{\partial f / \partial\beta_2}{\partial f/\partial x}, \frac{\partial f/ \partial b}{\partial f/\partial x}\right).$$
The asymptotic variance of radialization 50\% is thus
$$\mathbb{V} (\widehat{x}) \approx \left(\nabla_{\theta}x|_{\theta=\widehat{\theta}}\right)^TM(\xi)^{-1}\left(\nabla_{\theta}x|_{\theta=\widehat{\theta}}\right)$$
where $M(\xi)$ is the Fisher information matrix of $\theta$ associated with design $\xi$ evaluated at $\widehat{\theta}$. Further, if we are also interested in lethal dose 50\% (LD50) , and we want to estimate the ratio (suggested by Dr. Collins) $r=\frac{LD50}{RD50}$, then the asymptotic variance of $\widehat{r}$ can be approximated as
$$\mathbb{V}(\widehat{r})=\left(\begin{matrix}
    -\frac{\widehat{x}_{LD50}}{\widehat{x}_{RD50}^2} \\ \frac{1}{\widehat{x}_{RD50}}
\end{matrix}\right)^T\widehat{Cov}\left(\begin{matrix}
       \widehat{x}_{LD50}\\ \widehat{x}_{RD50}
    \end{matrix}\right)\left(\begin{matrix}
        -\frac{\widehat{x}_{LD50}}{\widehat{x}_{RD50}^2} \\\frac{1}{\widehat{x}_{RD50}}
    \end{matrix}\right) $$
    where the $2\times 2$ covariance matrix of $\widehat{x}_{LD50}$ and $\widehat{x}_{RD50}$ is computed by, again, the $\Delta$-method.
\end{example}

Plug-in $M(\xi)=\alpha s_0s_0^T+(1-\alpha) M_1$ where $s_0=\widetilde{s}(X)|_{x=0}$ and $M_1$ is the Fisher information matrix associated with the design with point $0$ removed. Let us consider $c$-optimality, i.e., $\Phi_c(M)=\log\mathbb{V}(\widehat{x})$, then by example~\ref{example:add_a_point}, we have
\begin{align}
    \Phi_c(M)&=\log\left[\frac{c^TM_1^{-1}c}{1-\alpha}-\frac{\alpha}{(1-\alpha)^2}c^TM_1^{-1}\widetilde{s}(X_0)\left(I+\frac{\alpha}{1-\alpha}\widetilde{s}(X_0)^TM_1^{-1}\widetilde{s}(X_0)\right)^{-1}\widetilde{s}(X_0)^TM_1^{-1}c\right]
\end{align}
where $c=\nabla_{\theta}x|_{\theta=\widehat{\theta}}$ is the gradient computed using the implicit function theorem. The Fréchet derivative of $\Phi_c$ at $M$ in the direction of $M_2$ is
\begin{align}\label{eq:equi_c}
F_{\Phi_c}(M,M_2)&=-\text{Tr}\left(cc^TM^{-1}\left(M_2-M\right)M^{-1}\right)\\
    &=c^TM^{-1}c-c^TM^{-1}\widetilde{s}(x)\widetilde{s}(x)^TM^{-1}c.
\end{align}
where $M$ is defined above and provided $M_2=\widetilde{s}(x)\widetilde{s}(x)^T$. Now we have all the necessary elements to derive the equivalence theorem for the two-stage robust dual-optimal design. A robust dual optimal design optimizes the following criterion 
\begin{align}\label{eq:robust-dual}
\Phi_{\text{robust-dual}}(M)=\frac{1}{K}\sum_{i=1}^K\left(\frac{\lambda_i}{p}\Phi_D(M|\widehat{\theta}_i)+(1-\lambda_i)\Phi_c(M|\widehat{\theta}_i)\right)
\end{align}
where $K$ is the number of sets of nominal values, $p$ is the dimension of $\theta_i$ $\lambda_i$ is the weight for $D-$ and $c-$optimality using the $i^{th}$ nominal values and $\widehat{\theta}_i$ represents the $i^{th}$ nominal values.
\begin{theorem}[Equivalence theorem for augmented two-stage robust dual-optimal design]\label{thm:equi_thm_two_stage} Suppose at first stage, we perform $K$ sets of experiments and derive $K$ sets of nominal values (denoted as $\widehat{\theta}_1,\cdots,\widehat{\theta}_K$). Given specific weights $\lambda_1,\cdots,\lambda_K$ and the proportion of added zero dose level $\alpha_i,i=1,\cdots,K$, suppose we are trying optimize the robust dual-optimal criterion \ref{eq:robust-dual}. Then a design $\xi$ is robust dual-optimal if and only if for any $x$, the  Fisher information associated with $\xi$ (denoted as $M$), satisfies
\begin{align}
    \frac{1}{K}\sum_{i=1}^K\left(\frac{\lambda_i}{p}F_{\Phi_D}(M,\widetilde{s}(x)\widetilde{s}(x)^T|\widehat{\theta}_i)+(1-\lambda_i)F_{\Phi_c}(M, \widetilde{s}(x)\widetilde{s}(x)^T|\widehat{\theta}_i)\right)\le0
\end{align}
where $F_{\Phi_D}(M,\widetilde{s}(x)\widetilde{s}(x)^T|\widehat{\theta}_i)$ is given by \ref{eq:equi_D} and $F_{\Phi_c}(M, \widetilde{s}(x)\widetilde{s}(x)^T|\widehat{\theta}_i)$ is given by \ref{eq:equi_c}.
\end{theorem}

\textcolor{black}{According to Theorem~\ref{thm:equi_thm_two_stage}, the researcher could apply a dual-criterion approach where weights are assigned to both D- and c-optimality objectives. The design that satisfies the conditions laid out in Theorem 3.4 would optimize this balance, ensuring robustness by minimizing variability across different possible parameter values for the drug's effect. By using a two-stage approach, the researcher can refine the design after an initial set of experiments, focusing on the most informative dose levels identified in the first stage.}

\textcolor{black}{Further, Theorem~\ref{thm:equi_thm_two_stage} guides the researcher in setting up the dual-optimal design that not only optimizes parameter estimation (D-optimality) but also ensures accurate dose-response relationship modeling (c-optimality) across various experimental conditions. This robustness is crucial in toxicology, where small inaccuracies in dose-response estimates can lead to significant errors in understanding the drug's safety and efficacy.}  

\begin{figure}[!htbp]
\centering
{\includegraphics[width=0.75\textwidth]{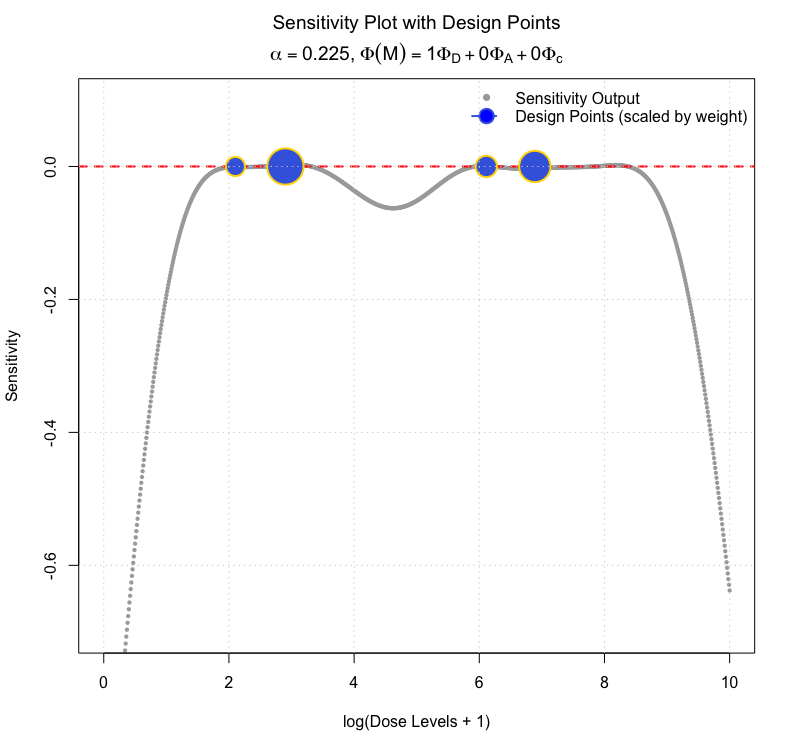}}
\caption{Sensitivity function of the sequential robust D-optimal design. \textcolor{black}{The x-axis represents the log-transformed dose levels, while the y-axis shows the sensitivity function values. The blue curve represents the sensitivity function across these dose levels, while the red line at the top of the plot highlights the zero-sensitivity baseline.}}\label{fig:robust_D_sens}
\end{figure}

\textbf{An Application}
\textcolor{black}{We apply Theorem~\ref{thm:equi_thm_two_stage} to the two-stage robust design scheme developed in Section~\ref{sec:meth}. Here we consider $\alpha=0.225$, i.e., 22.5\% of the second-stage observations will be put as 0 dose level and $22.5\%$ of the second-stage observations will be put as 10,000 dose level. Note that $0$ dose serves as the control group while $10,000$ dose level is to ensure that all sea urchins are dead, which is an important endpoint in toxicology. The sensitivity plot of the sequential D-optimal design is given in Figure~\ref{fig:robust_D_sens} and it verifies the global optimality of our implemented design despite of a little numerical fluctuation. We include more sensitivity plots under different optimality criterion and different $\alpha$-values in the Appendix~\ref{sec:appendix}.}

\section{Simulation Study: Sequential Optimal Designs for Bivariate Probit Model}\label{sec:simu}

 In this section, we develop two-stage optimal designs for bivariate probit model described in \cite{dragalin2008two}. Such a model has numerous applications in the context of Phase I or II studies, exploiting dose-finding that accounts for both efficacy and safety. 

 \subsection{The Model and the Fisher Information Matrix} Recently, various approaches have been suggested for dose escalation studies based on observations of both undesirable events and evidence of therapeutic benefit. bivariate probit model has been applied to model such a response and we give a brief introduction below; for a more comprehensive review, see \cite{mccullagh2019generalized}.

Let \( Y \in \{0, 1\} \) represent efficacy response and \( Z \in \{0, 1\} \) represent toxicity response, where 1 indicates occurrence and 0 indicates non-occurrence. In our example, the efficacy response is 'no VTE' and the toxicity response is 'bleeding'. A possible dose is denoted by \( x \). The probabilities of different response combinations are defined as:
\[
p_{yz}(x) = \text{Pr}(Y = y, Z = z \mid x), \quad y, z = 0, 1
\]
The probit model for correlated responses (see \cite{chib1998analysis, bekele2006dose}) is given by:
\[
p_{11}(x, \theta) = F(\theta_1^T f_1(x), \theta_2^T f_2(x), \rho) = \int_{-\infty}^{\theta_1^T f_1(x)} \int_{-\infty}^{\theta_2^T f_2(x)} \frac{1}{2\pi |\Sigma|^{1/2}} \exp\left( -\frac{1}{2} \mathbf{v}^T \Sigma^{-1} \mathbf{v} \right) dv_1 dv_2
\]
Here, \( \theta = (\theta_1^T, \theta_2^T)\), and the variance–covariance matrix is:
\[
\Sigma = \begin{pmatrix} 1 & \rho \\ \rho & 1 \end{pmatrix}
\]
The matrix \( \Sigma \) is assumed known, though similar results can be derived if \( \rho \) is unknown. The functions \( f_1(x) \) and \( f_2(x) \) include relevant covariates, such as \( f_1(x) = f_2(x) = (1, x)^T \) for modeling a single drug effect, or \( f_1(x) = f_2(x) = (1, x_1, x_2, x_1x_2)^T \) for modeling drug combinations with interaction effects \citep{ashford1970multi}. Additional covariates, such as age, weight, or dosage frequency, can also be incorporated into \( f_1 \) and \( f_2 \). The marginal distributions for the probabilities of efficacy \( p_{1.}(x, \theta) \) and toxicity \( p_{.1}(x, \theta) \) are:
\[
p_{1.}(x, \theta) = F(\theta_1^T f_1(x)) \quad \text{and} \quad p_{.1}(x, \theta) = F(\theta_2^T f_2(x))
\]
where
\[
F(v) = \int_{-\infty}^{v} \frac{1}{\sqrt{2\pi}} \exp\left(-\frac{u^2}{2}\right) du
\]
The other probabilities are derived as follows:
\[
p_{10}(x, \theta) = p_{1.}(x, \theta) - p_{11}(x, \theta)
\]
\[
p_{01}(x, \theta) = p_{.1}(x, \theta) - p_{11}(x, \theta)
\]
\[
p_{00}(x, \theta) = 1 - p_{1.}(x, \theta) - p_{.1}(x, \theta) + p_{11}(x, \theta)
\]
For clarity, we sometimes omit arguments like \( x \) and \( \theta \) when the context is clear.

\subsubsection{Information Matrix for Bivariate Probit Model}
The normalized and unnormalized Fisher information matrices are derived in detail in \cite{dragalin2008two} and we present their results here. For a more general procedure of deriving Fisher information for optimal designs, see \cite{atkinson2014elemental}.

Given a design
\begin{align*}
    \xi&=\left(\begin{matrix}
    x_1&x_2&\cdots& x_{n-1}&x_n\\
    p_1&p_2&\cdots& p_{n-1}&p_n
    \end{matrix}\right),
\end{align*}
 then the (normalized) Fisher information matrix under the bivariate probit model is
 $$M(\xi,\theta)=\sum_{i=1}^np_i\mu(x_i,\theta)$$
 where $\mu(x_i,\theta)$ is the elemental information for a single observation at dose $x$:
\[
\mu(x, \theta) = C_1 C_2 (P - pp^T)^{-1} C_2^T C_1^T
\]
with
\[
C_1 = \begin{pmatrix} \psi(\theta_1^T f_1) f_1 & 0 \\ 0 & \psi(\theta_2^T f_2) f_2 \end{pmatrix}, \quad C_2 = \begin{pmatrix} F(u_1) & 1 - F(u_1) & -F(u_1) \\ F(u_2) & -F(u_2) & 1 - F(u_2) \end{pmatrix}
\]
\[
u_1 = \frac{\theta_2^T f_2 - \rho \theta_1^T f_1}{\sqrt{1 - \rho^2}}, \quad u_2 = \frac{\theta_1^T f_1 - \rho \theta_2^T f_2}{\sqrt{1 - \rho^2}}
\]
\[
P = \begin{pmatrix} p_{11} & 0 & 0 \\ 0 & p_{10} & 0 \\ 0 & 0 & p_{01} \end{pmatrix} \quad \text{and} \quad p = (p_{11}, p_{10}, p_{01})^T
\]
where \( \psi(u) = \partial F(u)/\partial u \) denotes the probability density function of the standard normal distribution.

\subsection{$D$-optimality and $L$-optimality}

If the interest is purely in estimating the parameters $\theta=(\theta_1^T,\theta_2^T)$, then we can maximize the determinant of the total normalized information matrix $M(\xi,\theta)$. The resulting design is referred to as the $D$-optimal design. In practice, we may have specific quantities that are of great interest. For example, we may specify targeted probabilities $(p_{1.}^*,p_{.1}^*)$ in advance, representing the desired efficacy and toxicity levels. Suppose we want the response probabilities $(p_{1.}(x,\theta),p_{.1}(x,\theta))$ are closest to the targeted probabilities in the following sense:
\begin{align}\label{eq:optimization}
    d(\theta)=\min_x\left\{w[p_{1.}(x,\theta)-p_{1.}^*]^2+(1-w)[p_{.1}(x,\theta)-p_{.1}^*]^2\right\}
\end{align}
where $w$ is weight between $0$ and $1$. In other words, we are going to solve for the dose level $X^*$ such that the minimum in the above is obtained. Then the $L$-optimality is defined as
\begin{align}\label{eq:L-optimality}
    \Psi(M(\xi,\theta))=L^T(\theta)M^{-1}(\xi,\theta)L(\theta),\ L(\theta)=\frac{\partial X^*}{\partial\theta}.
\end{align}
It can also be viewed as $A$-criterion with $A=L(\theta)L^T(\theta)$ or $c$-optimality since $L(\theta)$ is a vector. Statistically speaking, $\Psi(M(\xi,\theta))$ represents the asymptotic variance of the estimator $X^*$. For locally optimal designs, we replace $\theta$ with its estimate $\widehat{\theta}$. As pointed out in \cite{dragalin2008two}, it is complicated to work with $p_{1.}$ and $p_{1.}$ directly; so we replace them with their quantiles, i.e., $\phi_1(x,\theta)=F^{-1}(p_{1.}(x))$, $\phi_2(x,\theta)=F^{-1}(p_{.1}(x))$, $\phi_1^*=F^{-1}(p_{1.}^*)$ and $\phi_2^*=F^{-1}(p_{.1}^*)$. \cite{dragalin2008two} shows that the minimizer in this case is 
$$X^*=\frac{(1-w)\theta_{22}(\phi_2^*-\theta_{21})+w\theta_{12}(\phi_1^*-\theta_{11})}{w\theta_{12}^2+(1-w)\theta_{22}^*}$$
and its partial derivative w.r.t. $\theta$ can be derived easily.

\subsection{Two Extensions}

\subsubsection{Extension with Penalty}

\cite{dragalin2008two} suggests to introduce penalty into $\Psi(M(\xi,\theta))$ for clinical trial considerations, see also \cite{haines2003bayesian,dragalin2006adaptive,dragalin2008adaptive}. Specifically, they define the total normalized penalty 
$$\Phi(\xi,\theta)=\int_{\mathcal{X}}\phi(x,\theta)\xi(dx)$$
where $\phi$ is a user-specified penalty function. For example, to control the costs associated with undesirable events like lack of efficacy or the occurrence of toxicity, then one can define
\begin{align}\label{eq:penalty_bivariate_probit}
    \phi(x,\theta)=p_{1.}^{-C_E}(1-p_{.1})^{-C_T}
\end{align}
where $C_E$ and $C_T$ are positive constants that quantify the relative importance of penalties for lack of efficacy and occurrence of toxicity, respectively. We use $\phi$ defined in Equation~\ref{eq:penalty_bivariate_probit} in the simulation studies (Section~\ref{sec:bivariate_probit_simu}). The resulting optimal design is 
\begin{align*}
    \xi^*&=\arg\min_{\xi\in\Xi}\Psi\left(\frac{M(\xi,\theta)}{\Phi(\xi,\theta)}\right).
\end{align*}
Note that the special choice $C_E=C_T=0$ corresponds to the un-penalized case.

\subsubsection{Extension to Two-stage Designs}\label{sec:ext_two_stage}
In practice, estimating \(\theta\) often relies on pilot data, leading to a two-stage design approach. This method is similar to the one outlined in Section \ref{sec:proposed_seq_robust_design}. Specifically, we begin by fixing a proportion \(\alpha \in [0,1]\) of the total pre-specified observations to be collected as pilot data. From this pilot data, we then obtain an estimate \(\widehat{\theta}\) using maximum likelihood estimation or other suitable methods. Based on \(\widehat{\theta}\), a locally optimal design is constructed according to a chosen criterion, such as \(D\)-optimality or \(L\)-optimality. The remaining \((1-\alpha)\) proportion of the observations is subsequently collected following this optimized design.

\subsection{Simulation Studies}\label{sec:bivariate_probit_simu}
In this subsection, we apply PSO to solve for both the $D$- and $L$-optimal designs in bivariate probit model and compare the results with those reported in \cite{dragalin2008two} where they used the first-order algorithm \citep{atkinson2007optimum,fedorov2012model}. Although the sensitivity plots fail to confirm the optimality of designs, PSO-generated designs beat the reported designs under most scenarios in terms of criterion values.

 \cite{dragalin2008two} suggests a nominal value $\widehat{\theta}=(-0.9,1.9,-3.98,3)$ obtained from a five-dose evenly spaced uniform design (such design is common when we do not have any prior knowledge):
$$\xi_u=\left(\begin{matrix}
    0.2&0.5&0.8&1.1&1.4\\
    0.2&0.2&0.2&0.2&0.2
\end{matrix}\right).$$
 Based on the estimated $\widehat{\theta}$, they also constructed and reported the $D$-optimal and $L$-optimal designs using first-order algorithms. For comparison, we take $\rho=0.5, w=0.5$ and $C_E=C_T=1$, consistent with the approach outlined in the paper. The results are presented in Table~\ref{tab:bivariate_probit} where we compare PSO-generated designs with the three benchmark designs reported in \cite{dragalin2008two}. As an illustration, consider the value \( 8.332 \) in row one, column one of the table. This value represents the criterion for the PSO-generated D-optimal one-stage design without penalty. Specifically, it corresponds to the one-stage D-optimality criterion without any penalties applied (i.e., \( C_E = C_T = 1 \)). This design is contrasted with the two-stage extension discussed in Section~\ref{sec:ext_two_stage}. Highlighted values in bold represent the best performance within each category, indicating the most efficient design strategy for minimizing D-optimality and L-optimality under specific conditions. 

 \textcolor{black}{The PSO approach used here is effective in finding designs that minimize specific criteria, making it an adaptive and efficient method to optimize both D-optimality and L-optimality. Key insights from this table include the advantages of PSO-generated designs, the impact of using one-stage versus two-stage designs, and the influence of applying penalties.}

 \textcolor{black}{Firstly, PSO-generated designs consistently outperform the benchmark designs in both one-stage and two-stage configurations, indicating that PSO effectively identifies optimal dose levels and weight distributions to improve statistical efficiency. Two-stage designs generally yield better performance than one-stage designs, as seen in lower values for both D-optimal and L-optimal criteria (e.g., 8.332 for one-stage D-optimality without penalty versus 8.413 for two-stage). This performance improvement with two-stage designs demonstrates the added flexibility and refinement they offer. Furthermore, applying penalties (i.e., adjusting \( C_E \) and \( C_T \) parameters) slightly increases the values of D-optimal and L-optimal criteria, highlighting a trade-off between strict optimality and practical constraints. Nevertheless, even with penalties, PSO-generated designs remain more efficient than the benchmark designs, underscoring the robustness of the PSO approach in real-world scenarios where constraints often exist.}

 \textcolor{black}{The superior performance of PSO-generated designs is evident across various design criteria. For example, PSO-generated designs achieve the lowest values in each category, such as 8.332 for one-stage D-optimality without penalty and 0.434 for L-optimality, confirming that they are highly efficient solutions. This consistency suggests that PSO is highly effective regardless of configuration—whether in one-stage, two-stage, penalized, or non-penalized setups. The PSO approach’s adaptability allows researchers to fine-tune experimental designs to meet specific needs, such as balancing dose-response optimization or addressing therapeutic study requirements. In practical terms, the ability of PSO-generated designs to minimize D-optimality and L-optimality with or without penalties positions it as a versatile tool for improving experimental design efficiency, especially in fields like clinical or pharmaceutical studies where optimal dose selection is crucial. These findings highlight PSO’s potential to enhance the quality of experimental data, reduce resource usage, and support more accurate conclusions in research.
}

\begin{table}[h!]
\centering
\rotatebox{90}{%
\begin{minipage}{\textheight} 
\centering
\caption{Comparison of $D$- and $L$-optimality Across One-stage and Two-stage Designs}
\begin{tabular}{lcccccccc}
\toprule
\multirow{3}{*}{\textbf{Designs}} & \multicolumn{4}{c}{\textbf{No Penalty}} & \multicolumn{4}{c}{\textbf{With Penalty}} \\ \cmidrule(lr){2-5} \cmidrule(lr){6-9}
& \multicolumn{2}{c}{\textbf{One-stage}} & \multicolumn{2}{c}{\textbf{Two-stage}} & \multicolumn{2}{c}{\textbf{One-stage}} & \multicolumn{2}{c}{\textbf{Two-stage}} \\ \cmidrule(lr){2-3} \cmidrule(lr){4-5} \cmidrule(lr){6-7} \cmidrule(lr){8-9}
& \textbf{D} & \textbf{L} & \textbf{D} & \textbf{L} & \textbf{D} & \textbf{L} & \textbf{D} & \textbf{L} \\ \midrule
\textbf{PSO-1 D-optimal [w/op]}      & \textbf{8.332}  & 1.025  & 8.413  & 0.868  & 33.166  & 499.573 & 33.167 & 423.126 \\ 
\textbf{PSO-1 L-optimal [w/op]}      & 13.820 & \textbf{0.434}  & 9.222  & 0.521  & 43.702  & 762.282 & 39.104 & 914.753 \\ 
\textbf{PSO-2 D-optimal [w/op]}      & 8.389  & 1.167  & \textbf{8.346}  & 0.921  & 37.448  & 1667.525 & 37.405 & 1317.194 \\ 
\textbf{PSO-2 L-optimal [w/op]}      & 16.245 & 0.648  & 9.392  & \textbf{0.506}  & 40.978  & 314.485 & 34.125 & 245.512 \\ 
\textbf{PSO-1 D-optimal [wp]}        & 10.763 & 3.623  & 8.538  & 1.100  & \textbf{15.290}  & 11.233  & 13.064 & 3.410 \\ 
\textbf{PSO-1 D-optimal [wp]}        & 9.576  & 1.555  & 8.449  & 0.807  & 16.710  & \textbf{9.253}   & 15.583 & 4.806 \\ 
\textbf{PSO-2 D-optimal [wp]}        & 16.730 & 89.155 & 8.867  & 1.581  & 19.544  & 180.210 & \textbf{11.133} & 3.196 \\ 
\textbf{PSO-2 L-optimal [wp]}        & 23.636 & 2994.772 & 8.785 & 1.607  & 25.984  & 5385.837 & 11.682 & \textbf{2.890} \\ 
\textbf{Reported D-optimal}                  & 8.654  & 0.680  & 8.551  & 0.682  & 31.193  & 190.555 & 31.090 & 191.143 \\ 
\textbf{Reported L-optimal}                  & 14.297 & 0.464  & 9.309  & 0.512  & 40.293  & 308.995 & 35.305 & 340.631 \\ 
\textbf{Uniform design}                      & 8.605  & 0.741  & 8.605  & 0.740  & 38.792  & 1403.310 & 38.792 & 1403.316 \\ 
\bottomrule
\end{tabular}
\caption*{\textit{Note: wp = with penalty, w/op = without penalty}}\label{tab:bivariate_probit}
\end{minipage}}
\end{table}

\subsection{Python Streamlit App}
Additionally, we have developed a Python Streamlit application, which is accessible online at \url{https://optimaldesignbivariateprobit.streamlit.app/}.

\section{Discussion}
In this chapter, we have designed a new sequential design scheme for toxicologists to design their experiments more efficiently and more robust in estimating dose-response relationships as well as important endpoints they are interested in. We provide equivalence theorems for checking the optimality of the design in terms of estimating the RD50 and volume of the confidence ellipsoid. Further, based on the four datasets that Dr. Collins has collected, we empirically show that
\begin{itemize}
    \item The zero-dose level is crucial in either estimating endpoint or dose-response curves while the extremely high dose level is necessary when we want to estimate the dose-response curve. In conventional statistical literature, the added practical dose levels are always neglected by statisticians;
    \item The two-stage designs are robust in the way they are constructed. For instance, the two-stage robust $D$-optimal has relatively high $D$-efficiencies under different sets of estimated parameters.
\end{itemize}

\bibliographystyle{agsm}

\bibliography{Bibliography-MM-MC,add_ref}

\newpage
\section{Appendix}\label{sec:appendix}
\subsection{Some Theoretical Developments}
Here we derive the equivalence theorem for other optimalities. Suppose
$$\Phi_A(M)=Tr\left( M^{-1}\right)=Tr\left( (\alpha s_0s_0^T+(1-\alpha)A_1)^{-1}\right).$$
Then its Fréchet derivative at $M_1$ in the direction of $M_2$ is
\begin{align*}
    F_{\Phi_A}(M_1,M_2)&=-(1-\alpha)Tr\left(M^{-1}(M_2-M_1)M^{-1}\right).
\end{align*}
The above Fréchet derivative coincides with those in conventional optimal design literature if we set $\alpha=0$. Similar sensitivity function for a multiple objective optimal design can be derived similarly. For instance, set
$$\Phi(M)=\frac{1}{K}\sum_{i=1}^K\left(\lambda_i^1\\Phi_D^i+\lambda_i^2\Phi_A^i+(1-\lambda_i^1-\lambda_i^2)\Phi_c^i\right)$$
where $K$ is the number of sets of nominal values and $i$ is to emphasize the dependency of locally optimal design. The resulting design $M$ is optimally optimal if the following holds for all $x$:
\begin{align}
    \sum_{i=1}^K\left(\frac{\lambda_i^1}{p}F_{\Phi_D}^i+\lambda_i^2F_{\Phi_A}^i+(1-\lambda_i^1-\lambda_i^2)F_{\Phi_c}\right)\le 0.
\end{align}
We next give some simulations on the above equivalence theorem and sensitivity functions.

\subsection{Sensitivity Plots and Multiple Optimality}
\subsubsection{D-optimality}
    \begin{figure}[h!]
        \centering
        \includegraphics[width=0.3\textwidth]{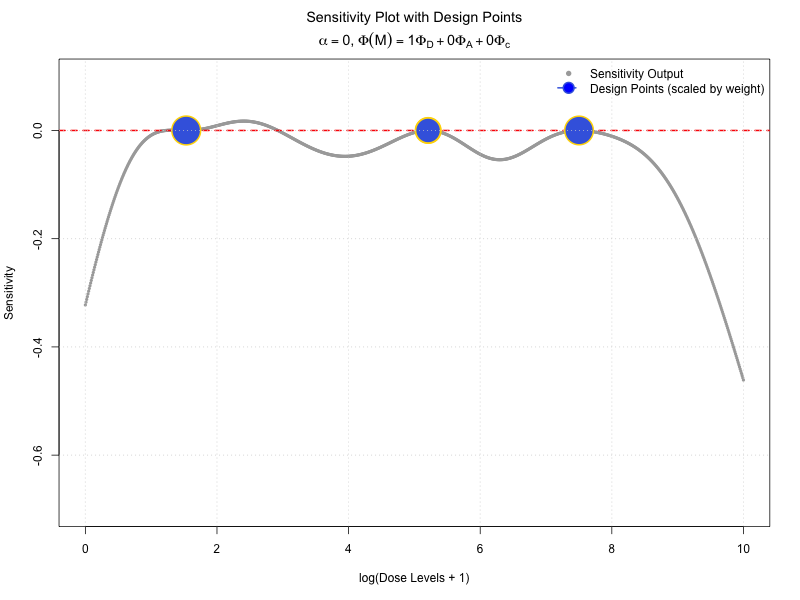}
        \includegraphics[width=0.3\textwidth]{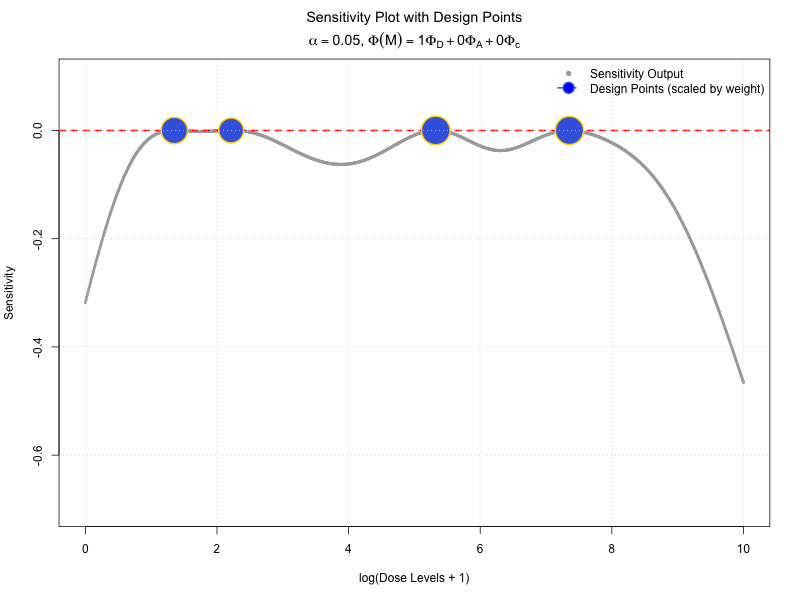}
        \includegraphics[width=0.3\textwidth]{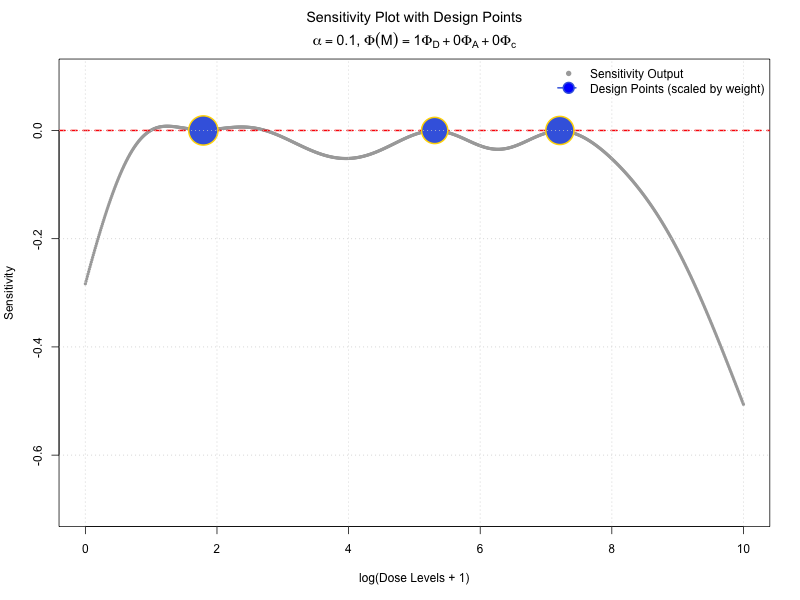}\\
        \includegraphics[width=0.3\textwidth]{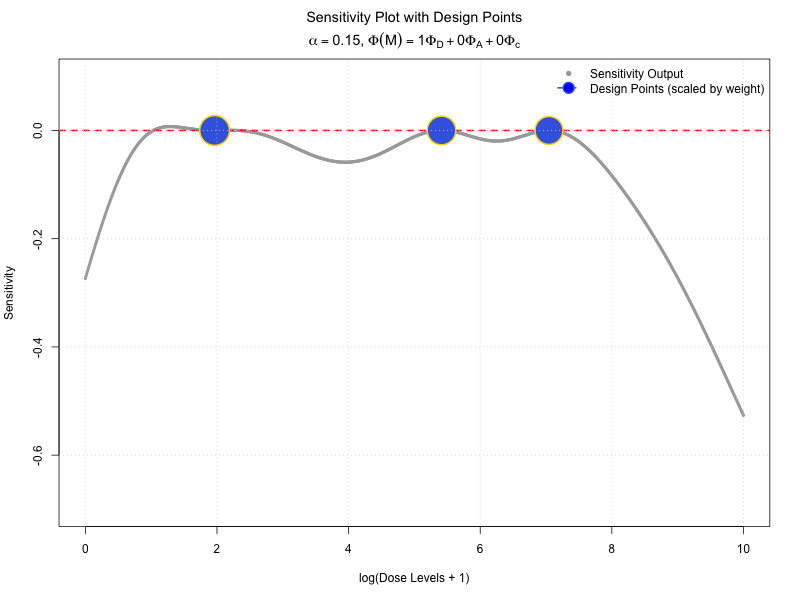}
        \includegraphics[width=0.3\textwidth]{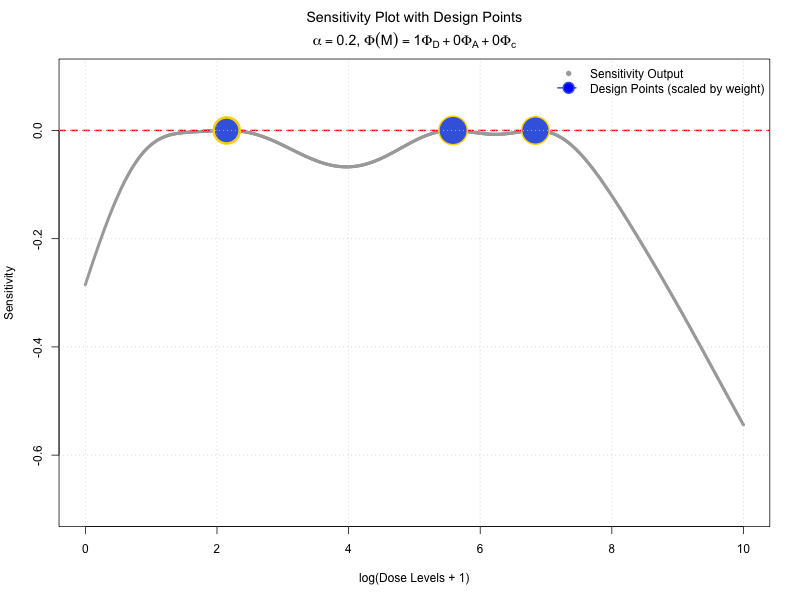}
        \includegraphics[width=0.3\textwidth]{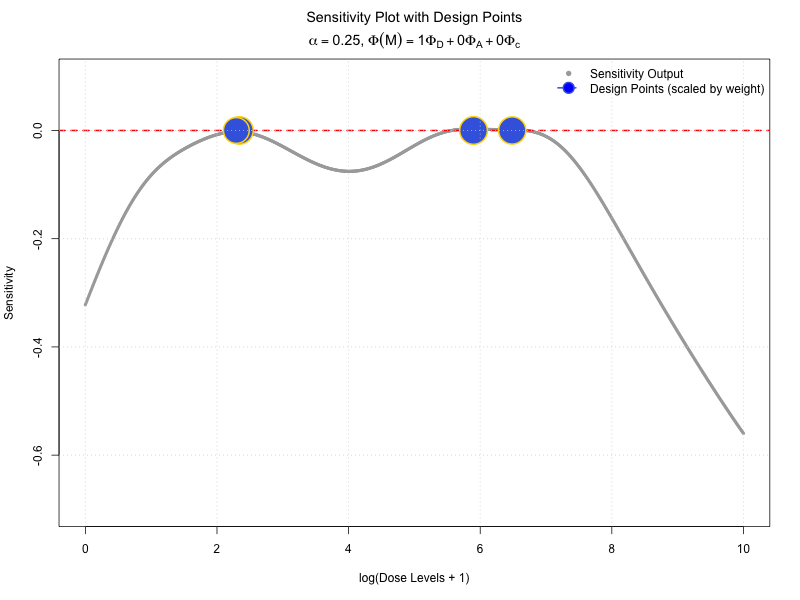}\\
   \includegraphics[width=0.3\textwidth]{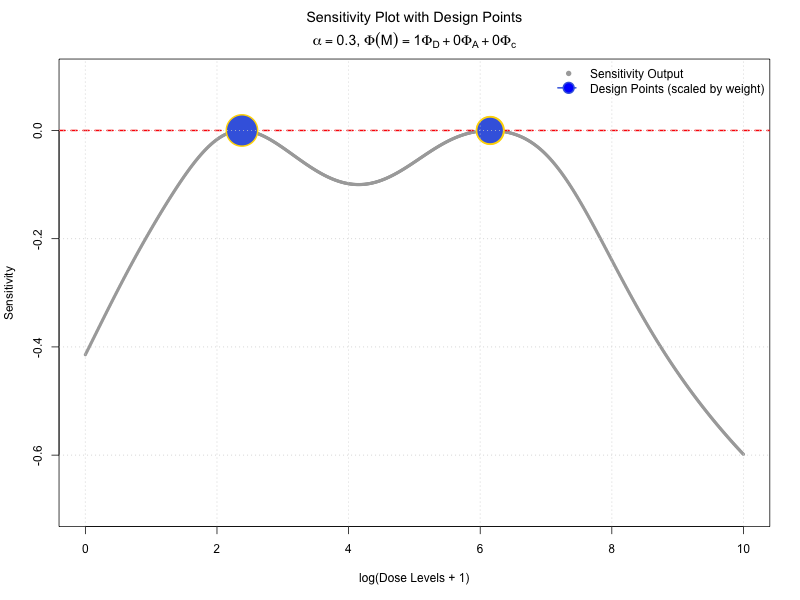}
        \includegraphics[width=0.3\textwidth]{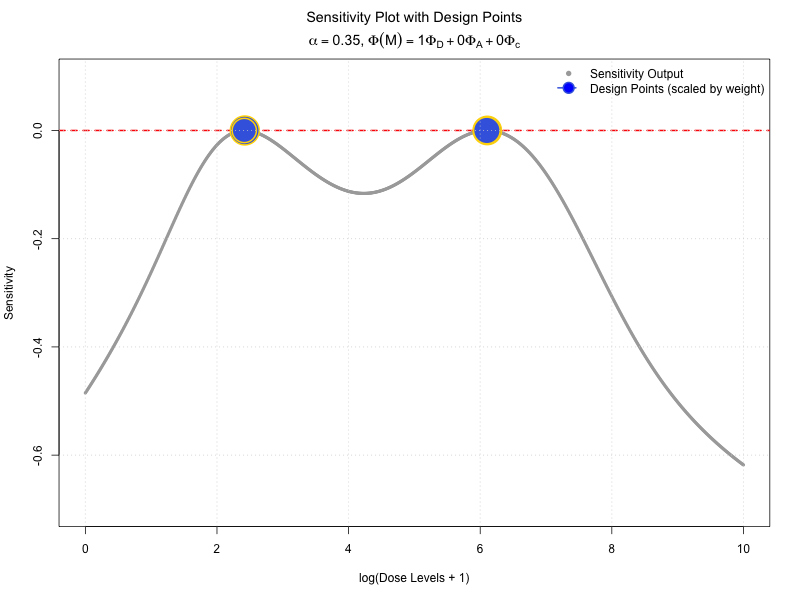} \includegraphics[width=0.3\textwidth]{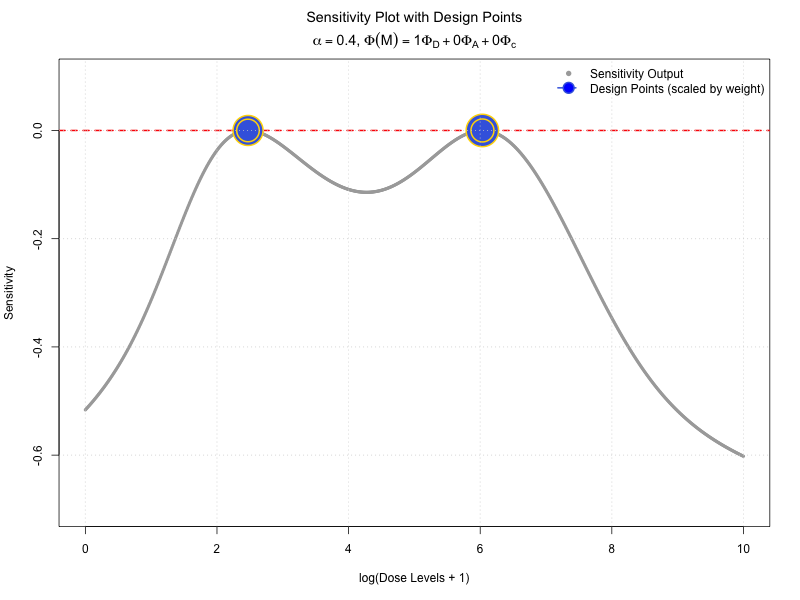}  \\      \includegraphics[width=0.3\textwidth]{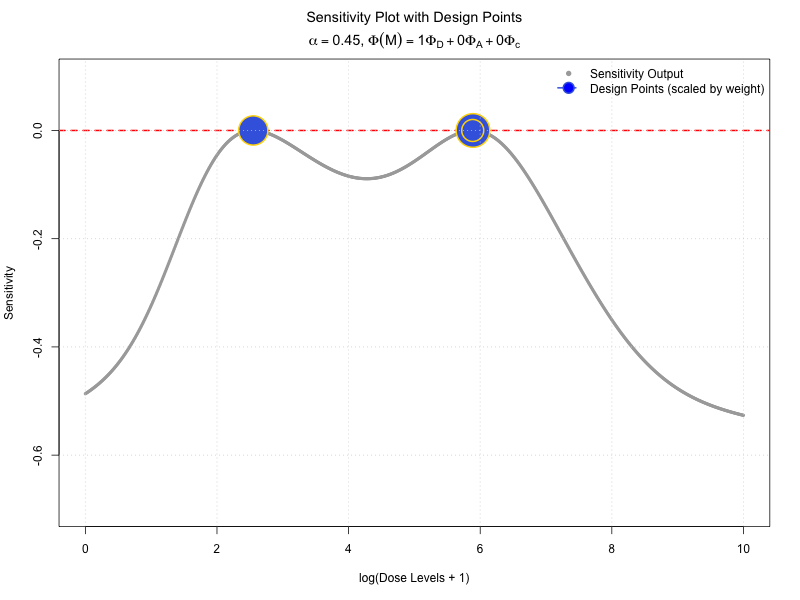}
        \caption{Sensitivity plots for D-optimality.}
    \end{figure}

    \subsubsection{DA-optimality with equal weights}
    \begin{figure}[h!]
        \centering
        \includegraphics[width=0.3\textwidth]{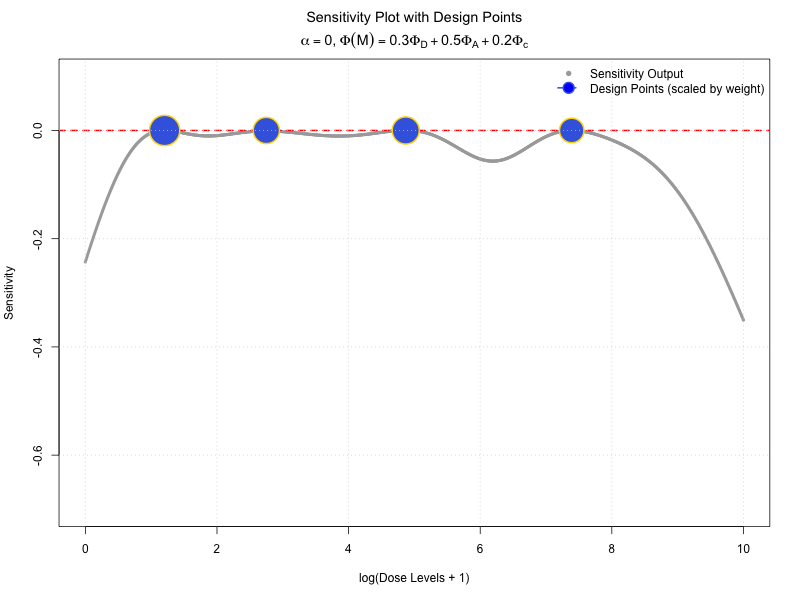}
        \includegraphics[width=0.3\textwidth]{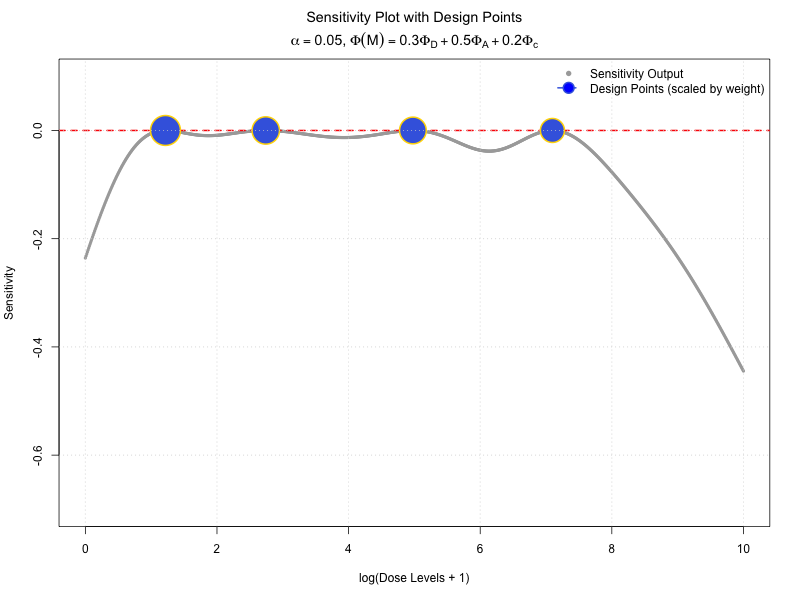}
        \includegraphics[width=0.3\textwidth]{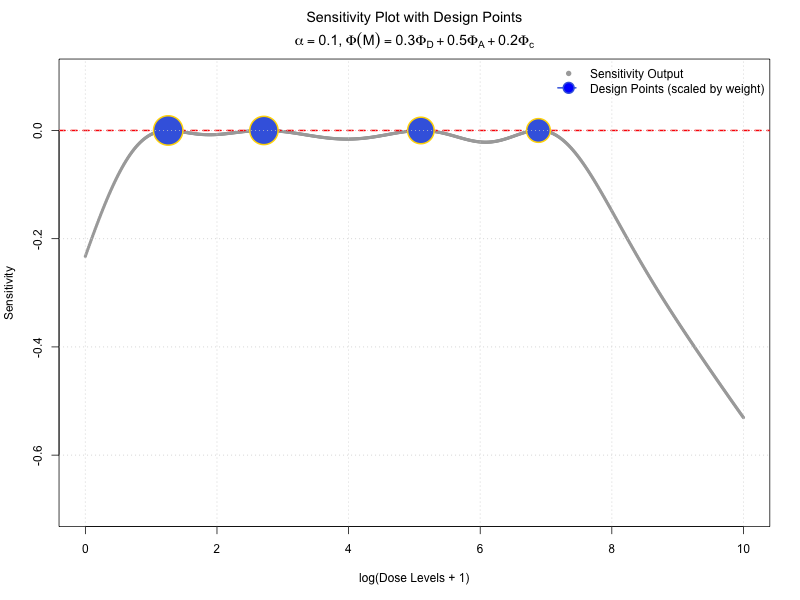}\\
        \includegraphics[width=0.3\textwidth]{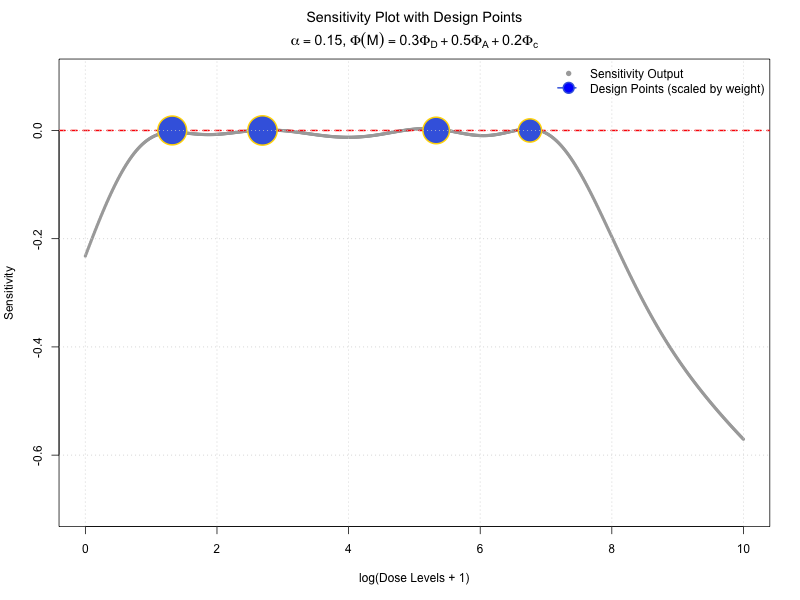}
        \includegraphics[width=0.3\textwidth]{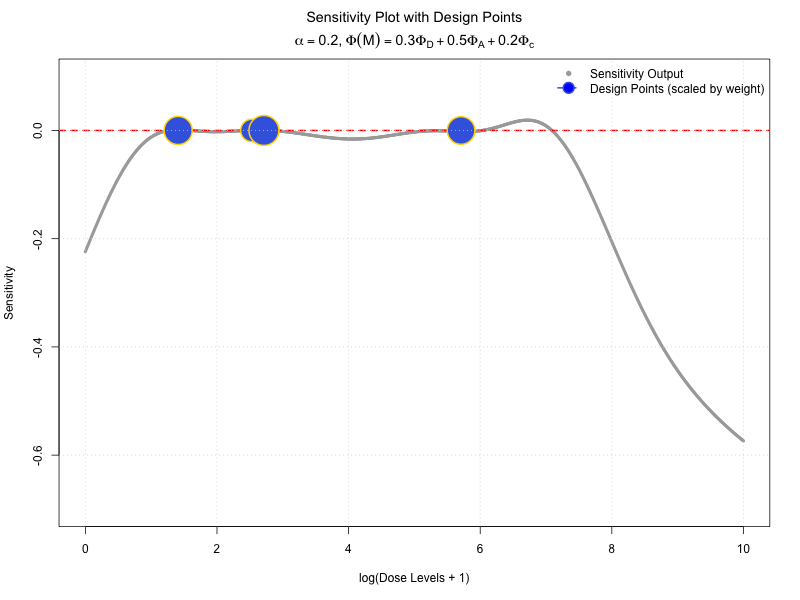}
        \includegraphics[width=0.3\textwidth]{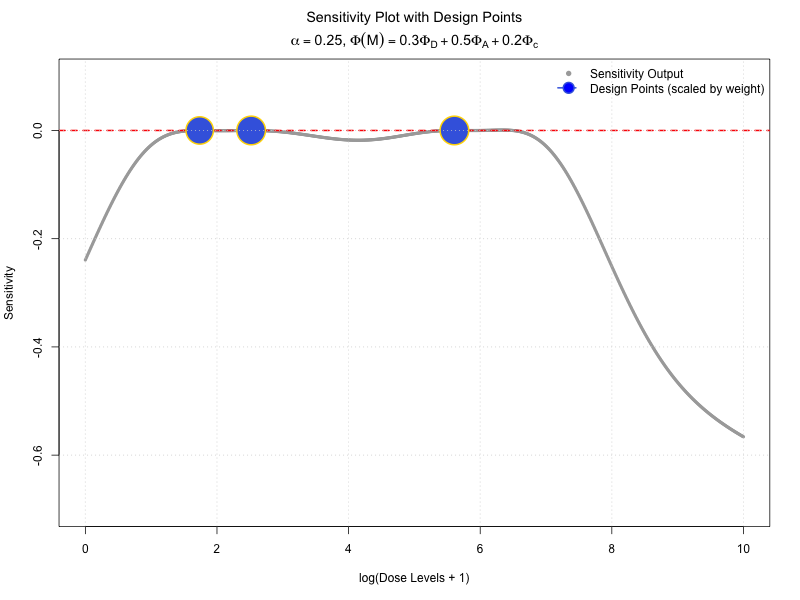}\\
   \includegraphics[width=0.3\textwidth]{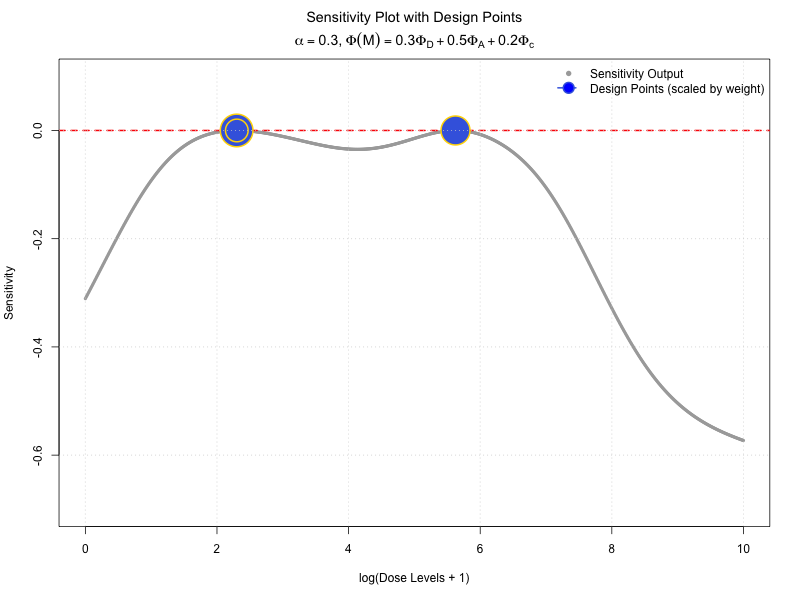}
        \includegraphics[width=0.3\textwidth]{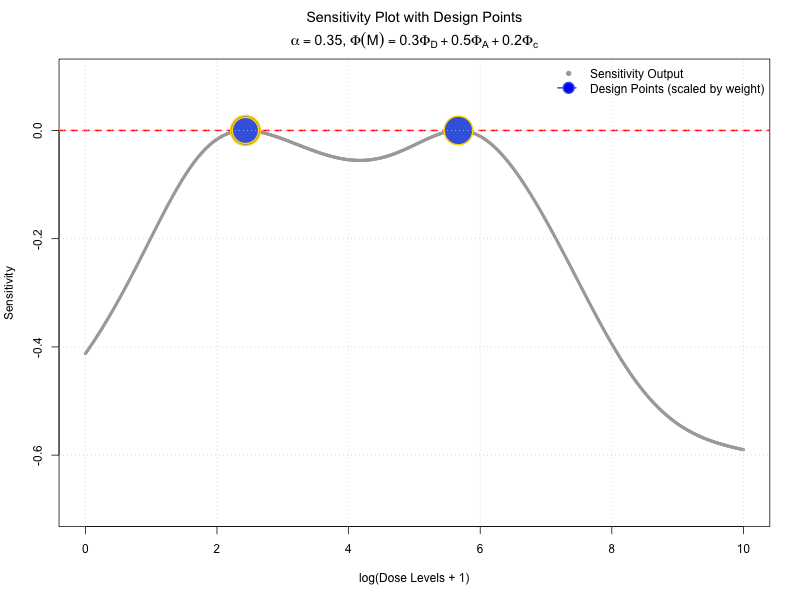} \includegraphics[width=0.3\textwidth]{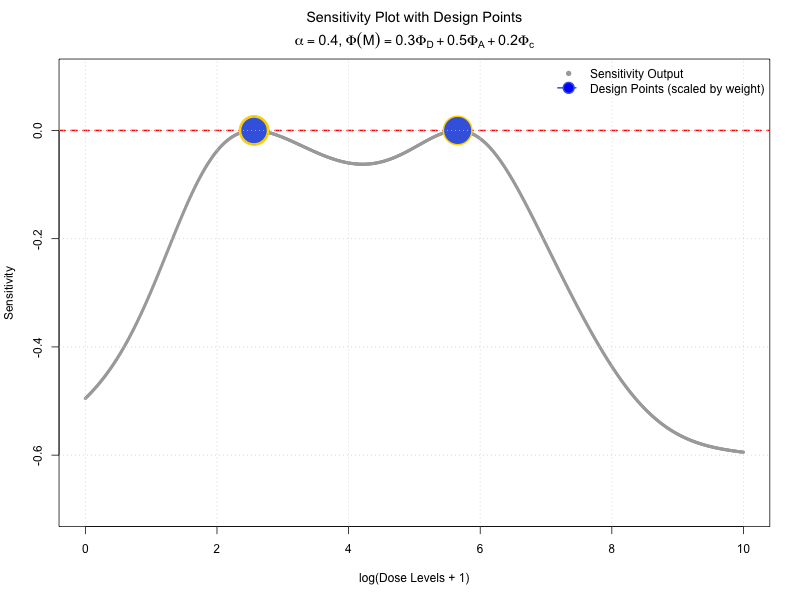}  \\      \includegraphics[width=0.3\textwidth]{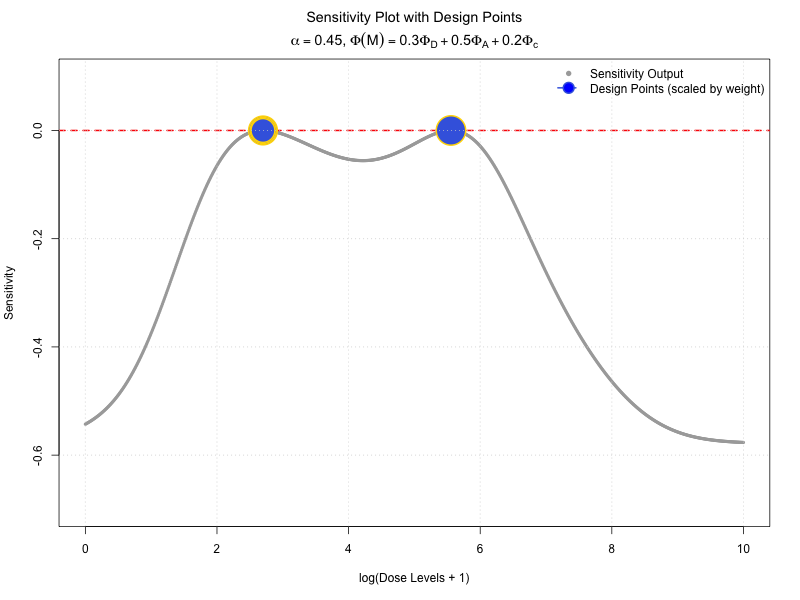}
        \caption{Sensitivity plots for DA-optimality.}
    \end{figure}

\subsubsection{Dc-optimality}
    \begin{figure}[h!]
        \centering
        \includegraphics[width=0.3\textwidth]{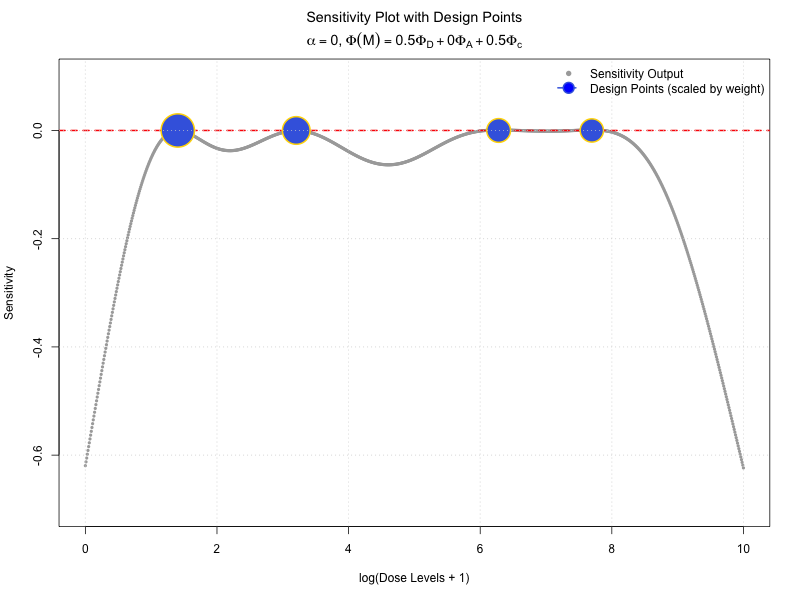}
        \includegraphics[width=0.3\textwidth]{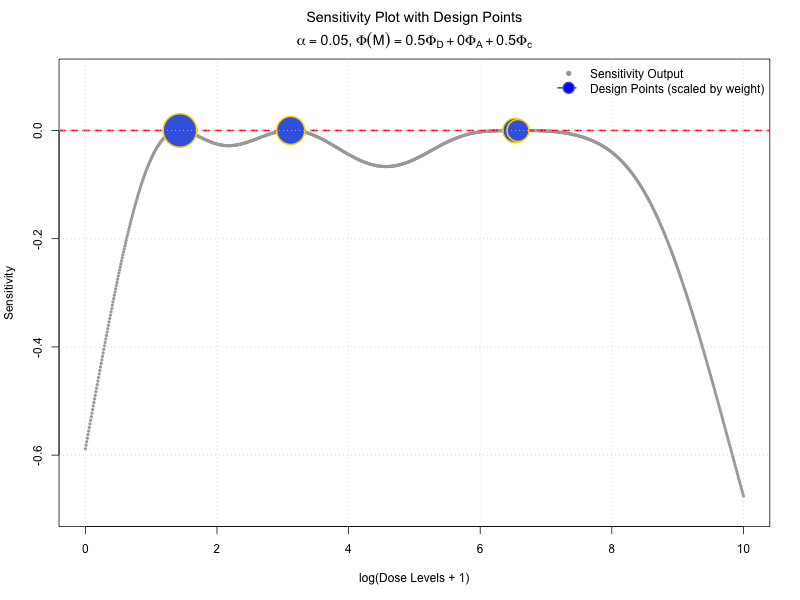}
        \includegraphics[width=0.3\textwidth]{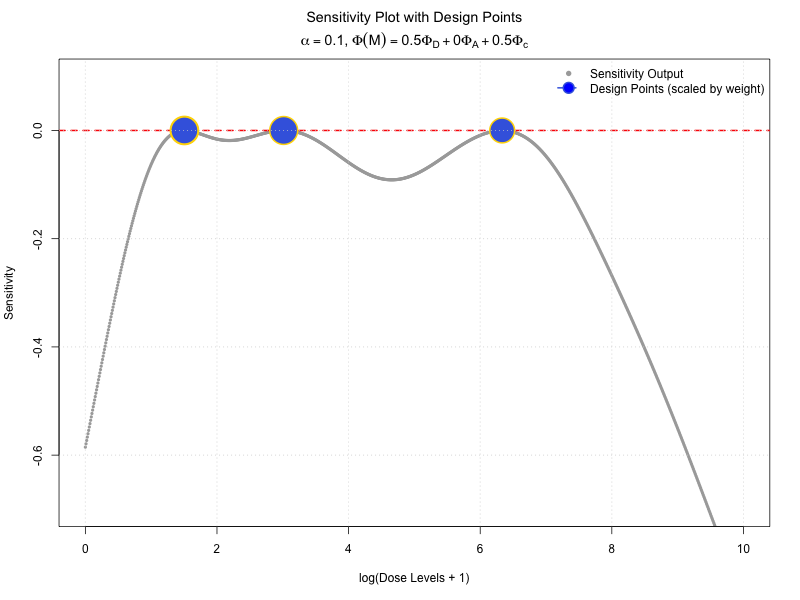}\\
        \includegraphics[width=0.3\textwidth]{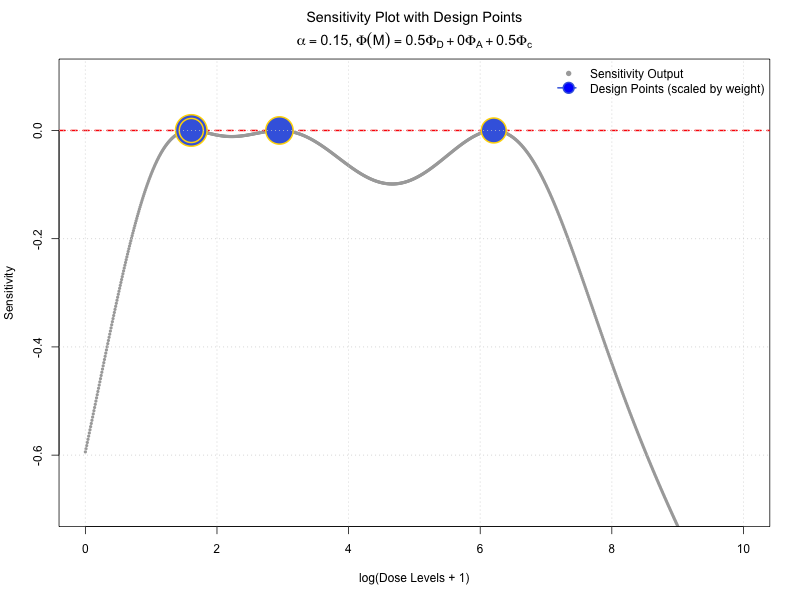}
        \includegraphics[width=0.3\textwidth]{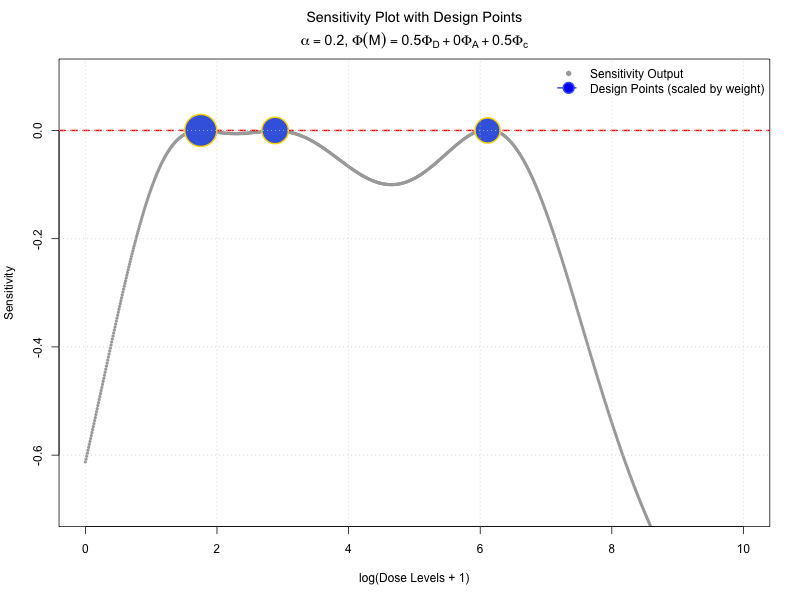}
        \includegraphics[width=0.3\textwidth]{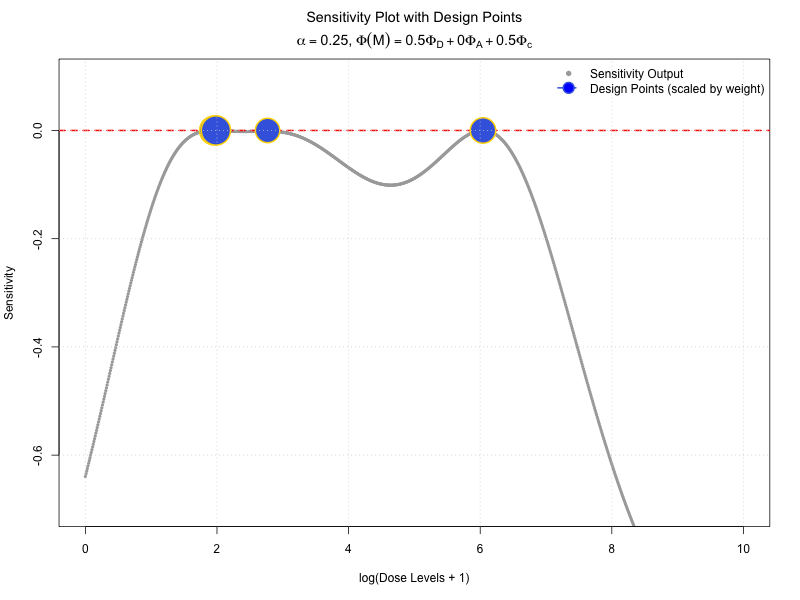}\\
   \includegraphics[width=0.3\textwidth]{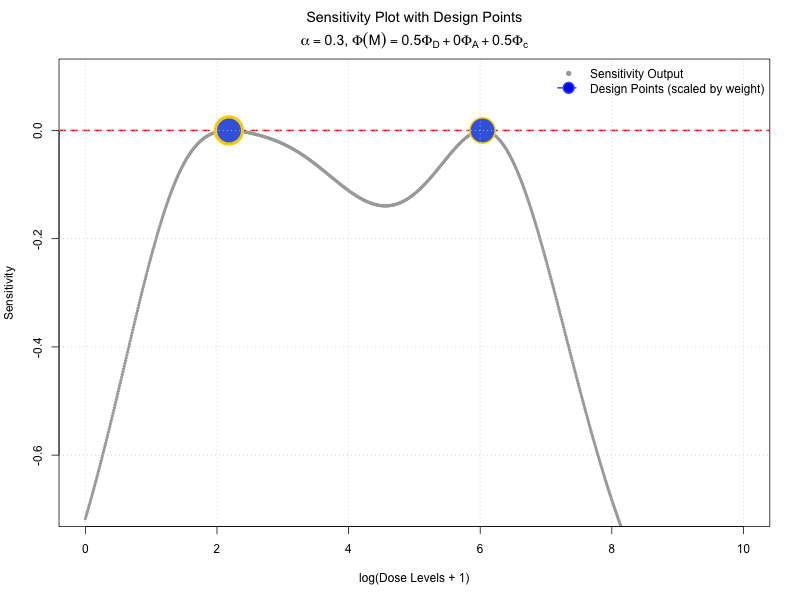}
        \includegraphics[width=0.3\textwidth]{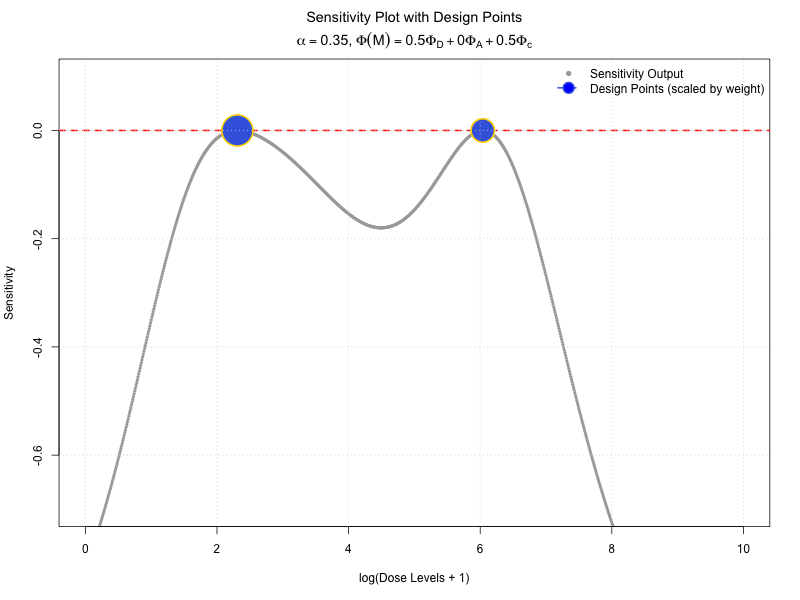} \includegraphics[width=0.3\textwidth]{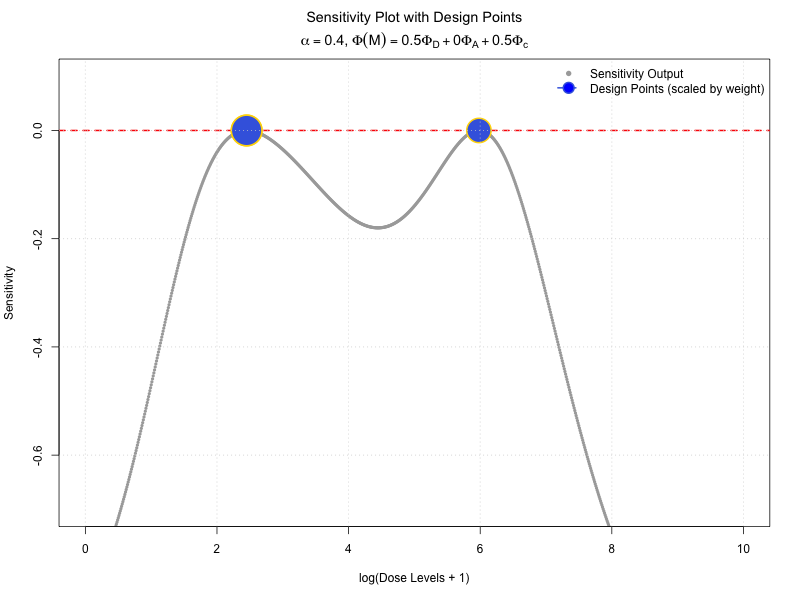}  \\      \includegraphics[width=0.3\textwidth]{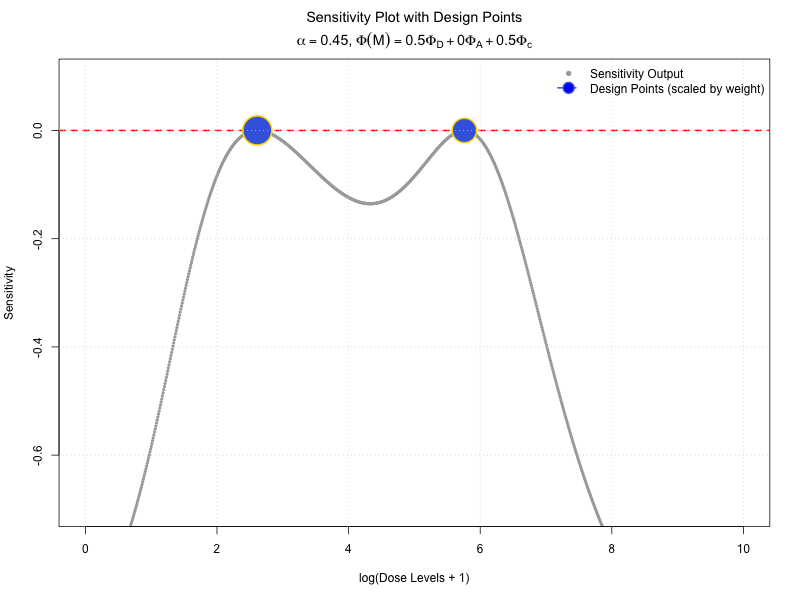}
        \caption{Sensitivity plots for Dc-optimality.}
    \end{figure}

\subsubsection{Ac-optimality}
    \begin{figure}[h!]
        \centering
        \includegraphics[width=0.3\textwidth]{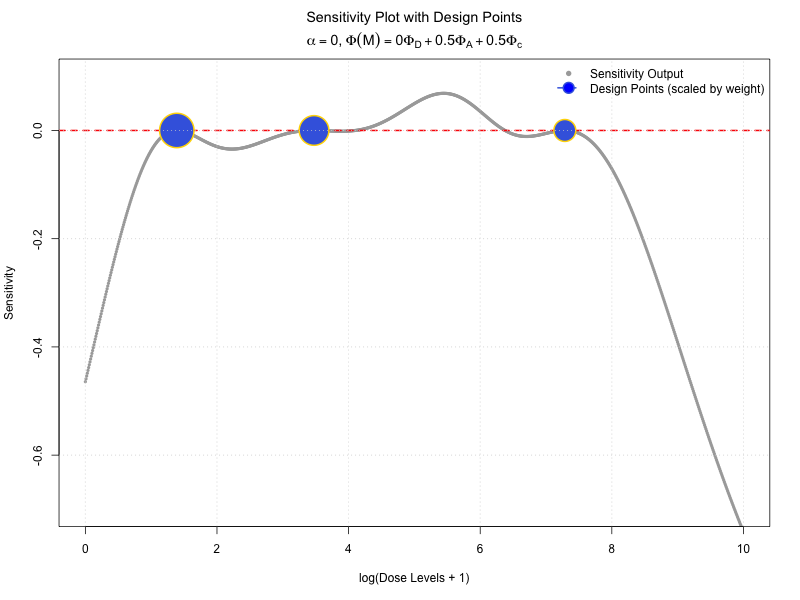}
        \includegraphics[width=0.3\textwidth]{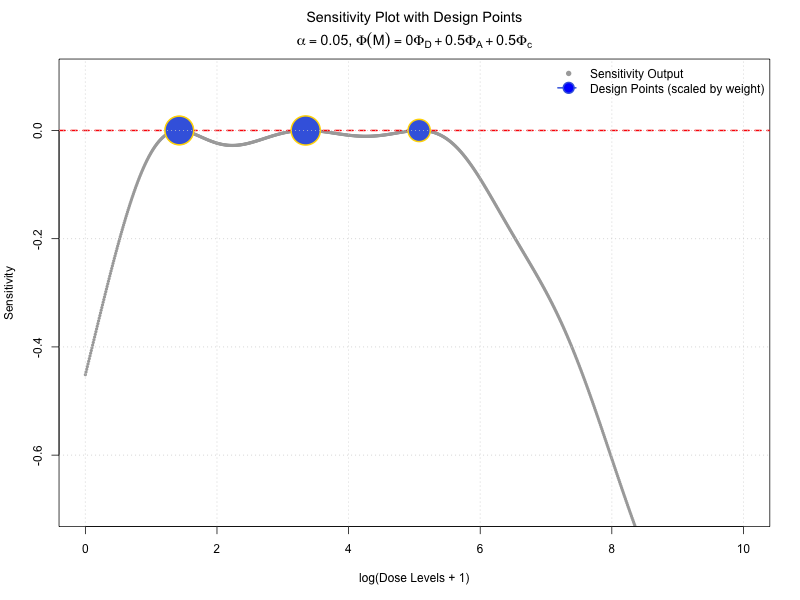}
        \includegraphics[width=0.3\textwidth]{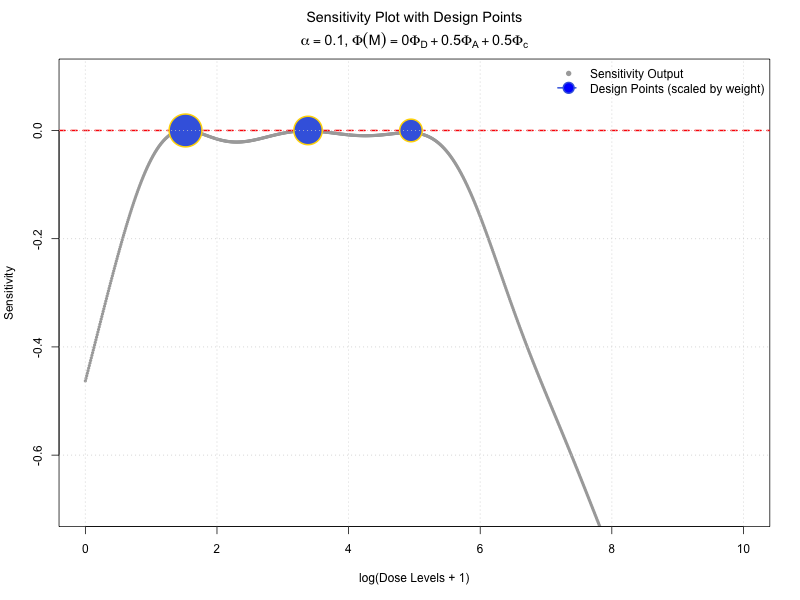}\\
        \includegraphics[width=0.3\textwidth]{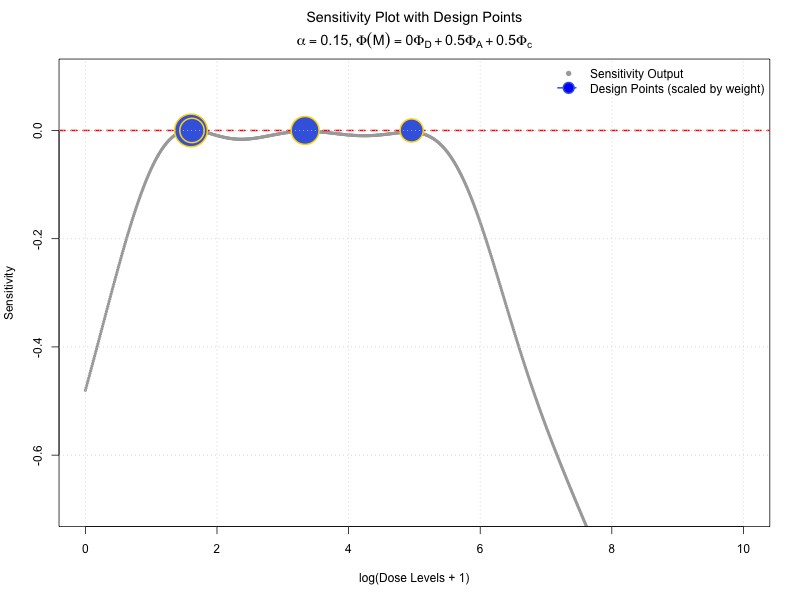}
        \includegraphics[width=0.3\textwidth]{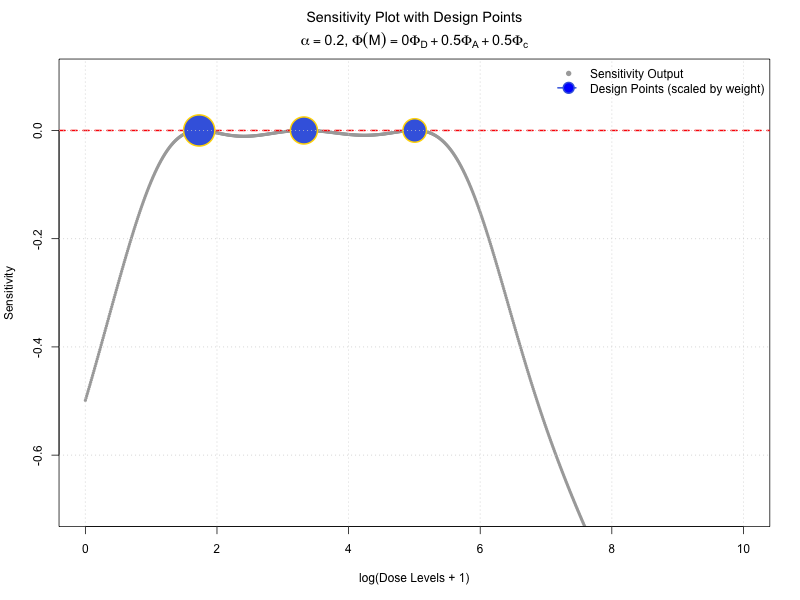}
        \includegraphics[width=0.3\textwidth]{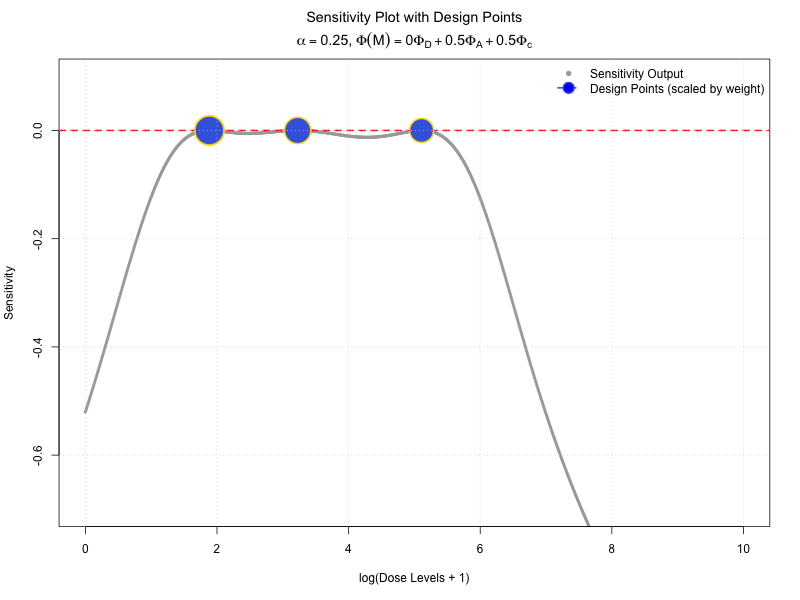}\\
   \includegraphics[width=0.3\textwidth]{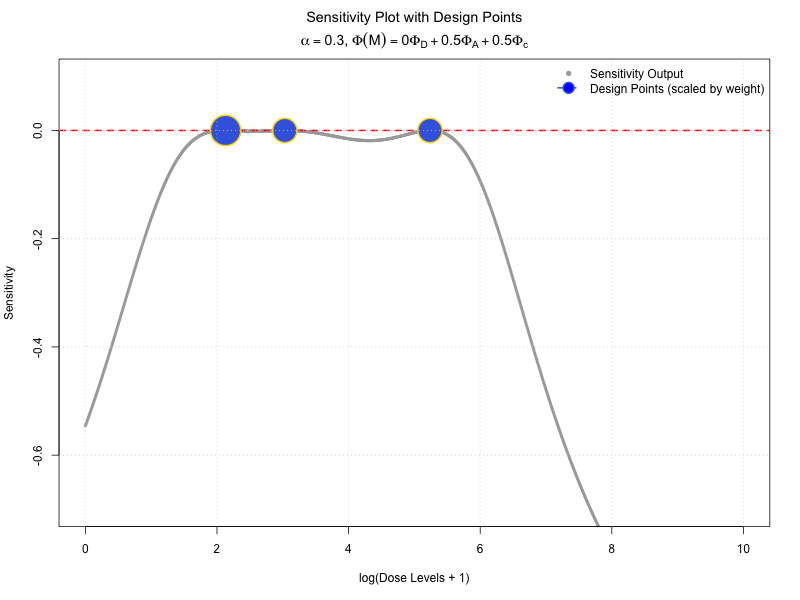}
        \includegraphics[width=0.3\textwidth]{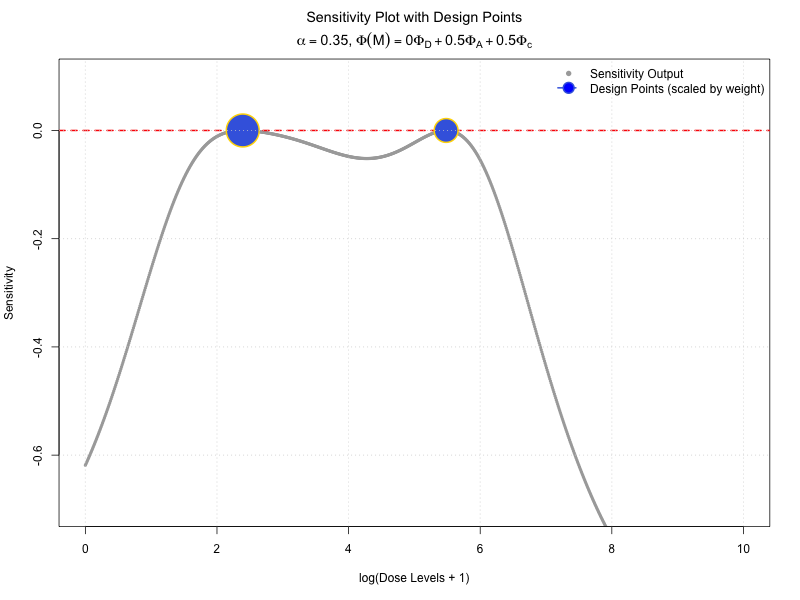} \includegraphics[width=0.3\textwidth]{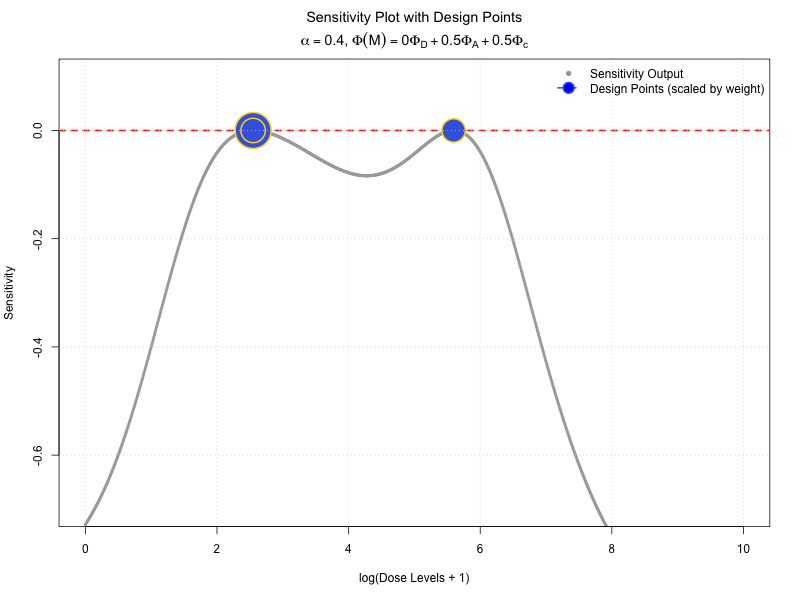}  \\      \includegraphics[width=0.3\textwidth]{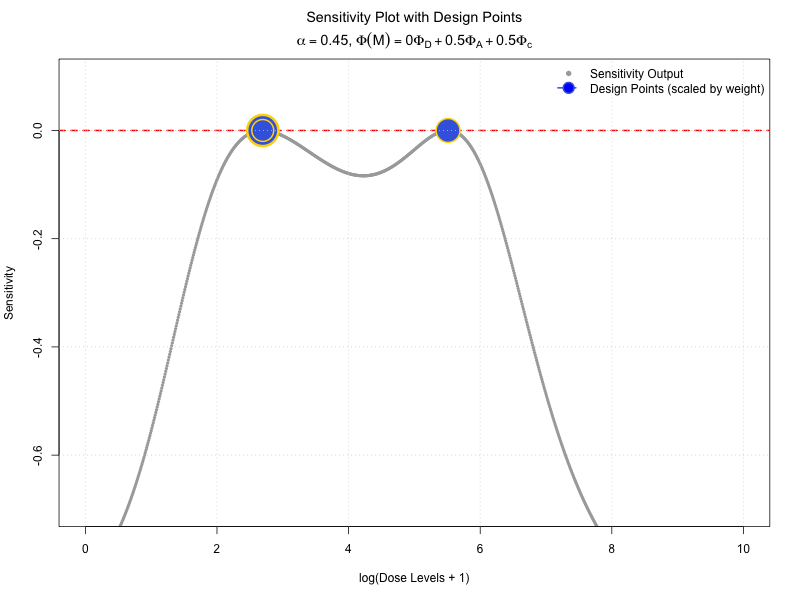}
        \caption{Sensitivity plots for Ac-optimality.}
    \end{figure}

\subsubsection{Multiple-optimality}
    \begin{figure}[h!]
        \centering
        \includegraphics[width=0.3\textwidth]{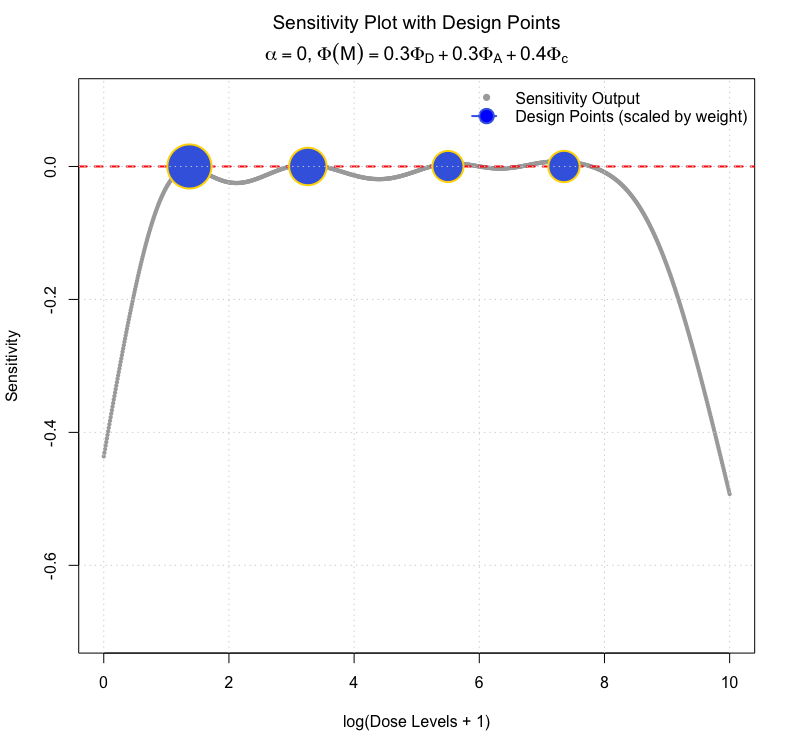}
        \includegraphics[width=0.3\textwidth]{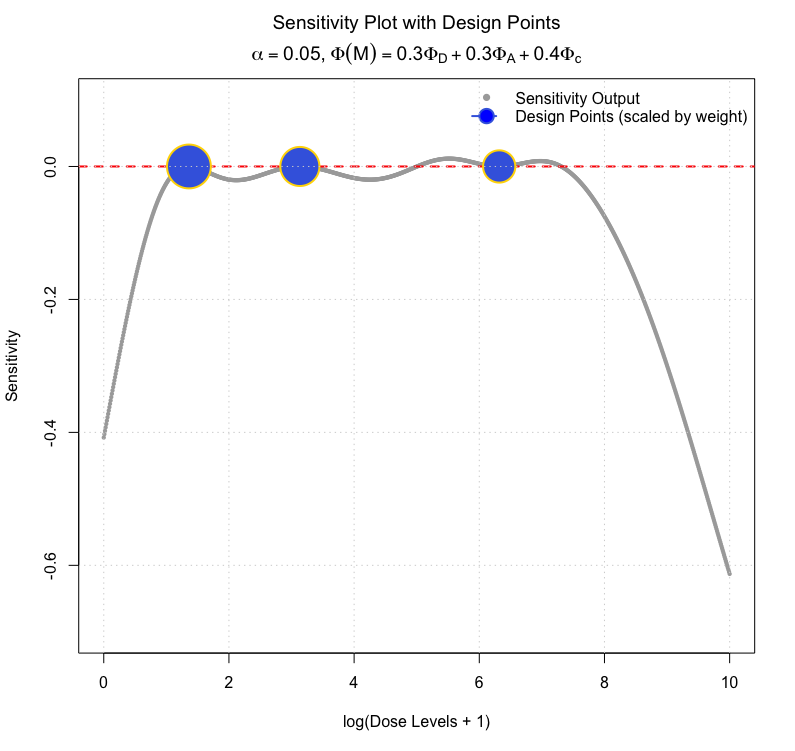}
        \includegraphics[width=0.3\textwidth]{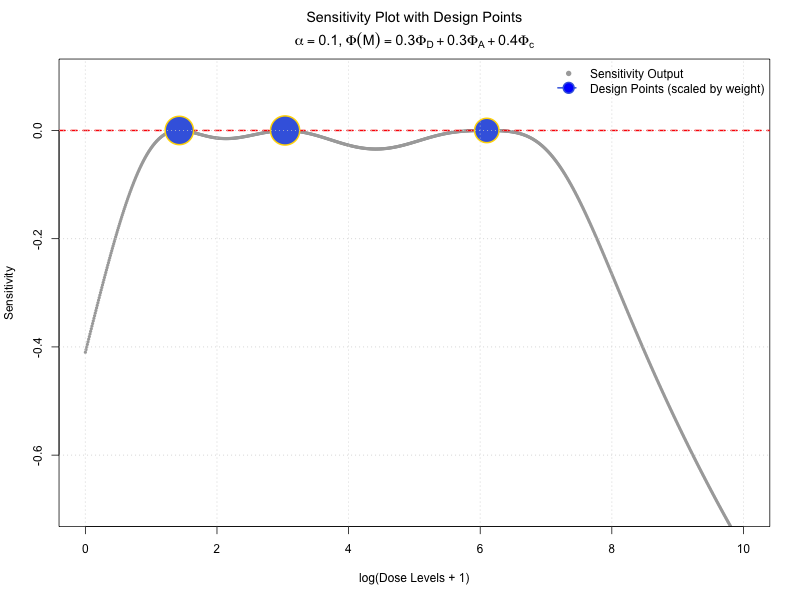}\\
        \includegraphics[width=0.3\textwidth]{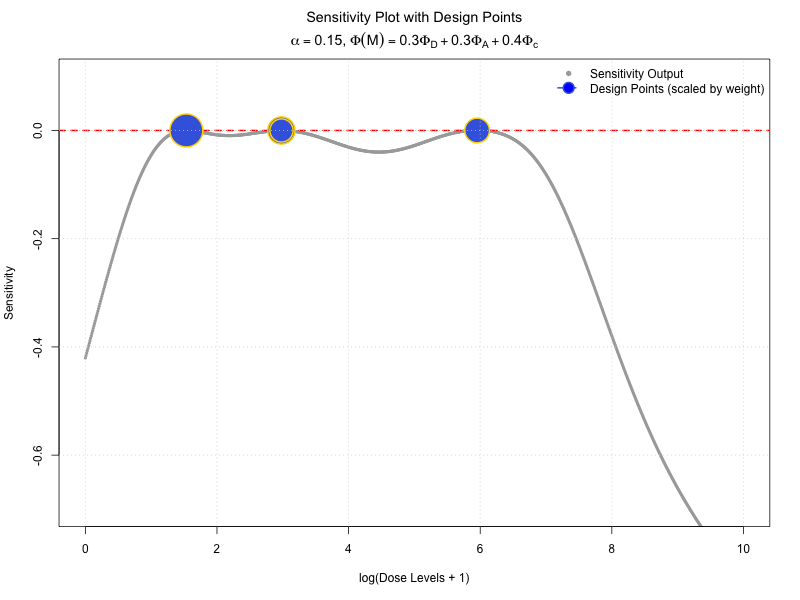}
        \includegraphics[width=0.3\textwidth]{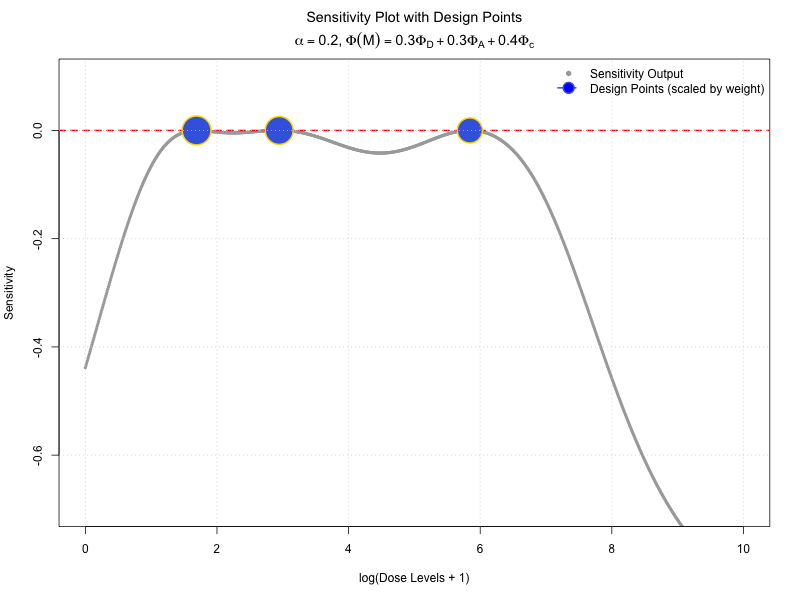}
        \includegraphics[width=0.3\textwidth]{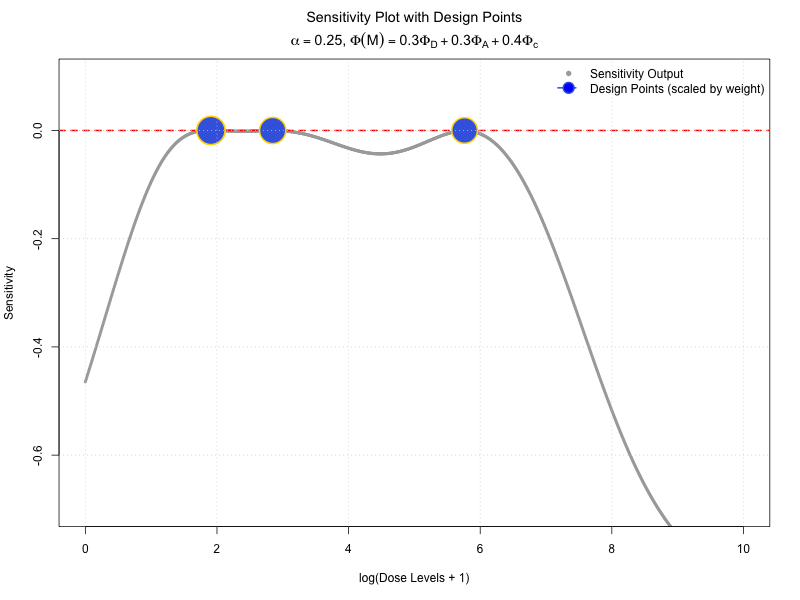}\\
   \includegraphics[width=0.3\textwidth]{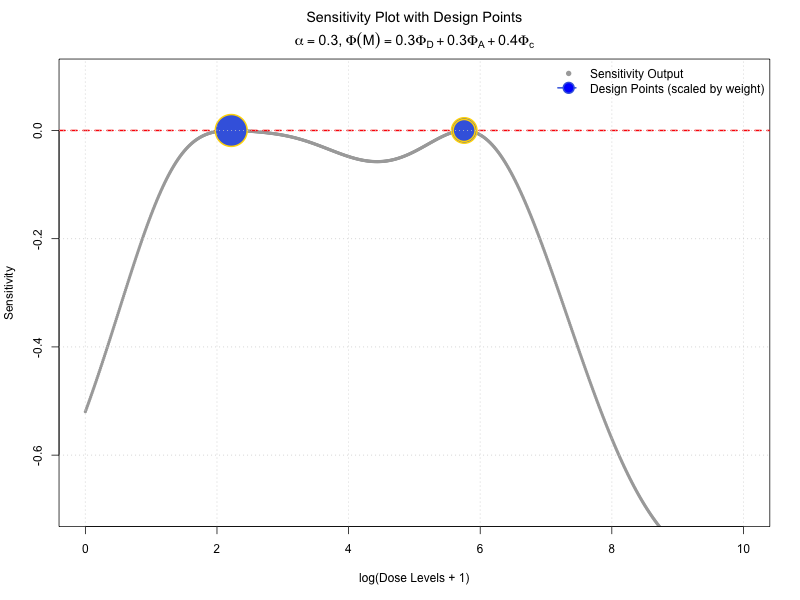}
        \includegraphics[width=0.3\textwidth]{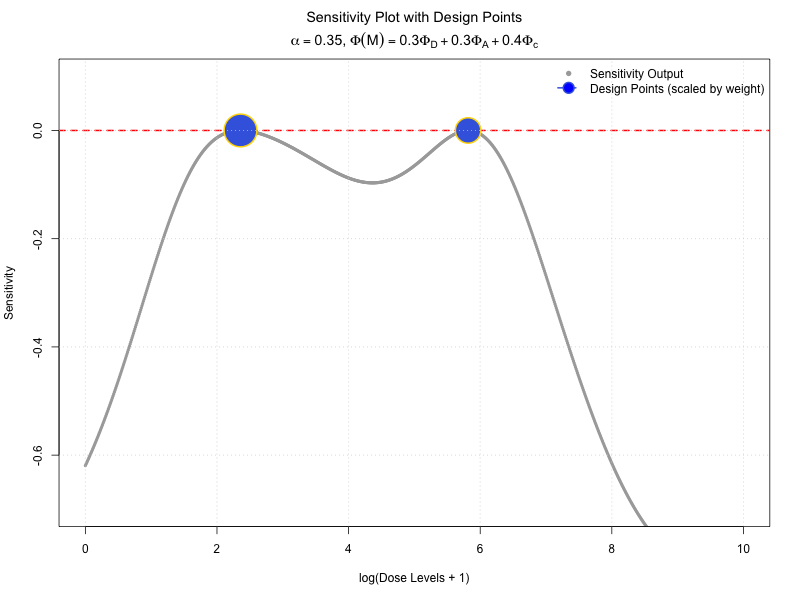} \includegraphics[width=0.3\textwidth]{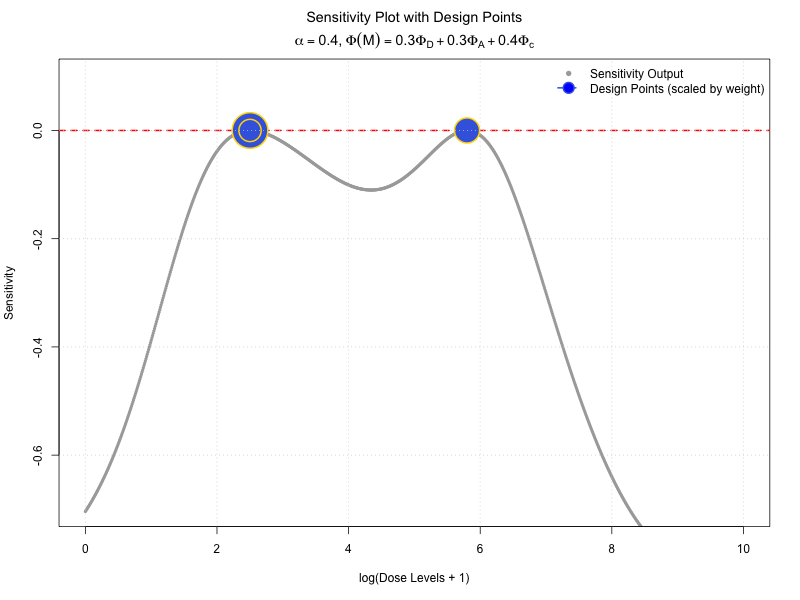}  \\      \includegraphics[width=0.3\textwidth]{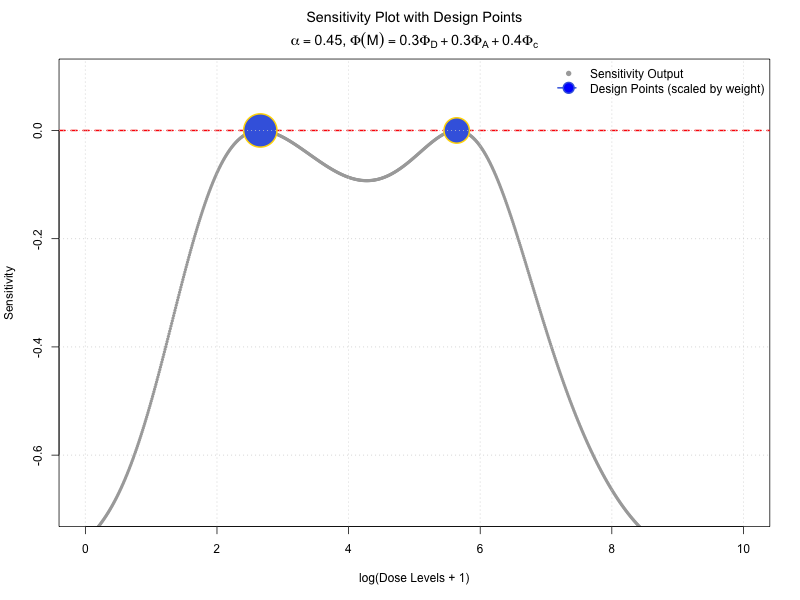}
        \caption{Sensitivity plots for Multiple-optimality.}
    \end{figure}

\newpage
    \subsubsection{A Nine-parameter Model}
    Finally, we plot the sensitivity function of a nine-parameter model of the following form:
    \begin{align*}
    \eta_1&=\log\left(\frac{\pi_1}{\pi_2+\pi_3}\right)=\beta_1+\alpha_1 x+\gamma_1x^2+\tau_1\sin(2x)\\
    \eta_2&=\log\left(\frac{\pi_1+\pi_2}{\pi_3}\right)=\beta_2+\alpha_2 x+\gamma_2x^2+\tau_2\sin(2x)\\
    \eta_3&=\log(\pi_1+\pi_2+\pi_3)=0
\end{align*}

    \begin{figure}[htbp!]
        \centering
        \includegraphics[width=0.3\textwidth]{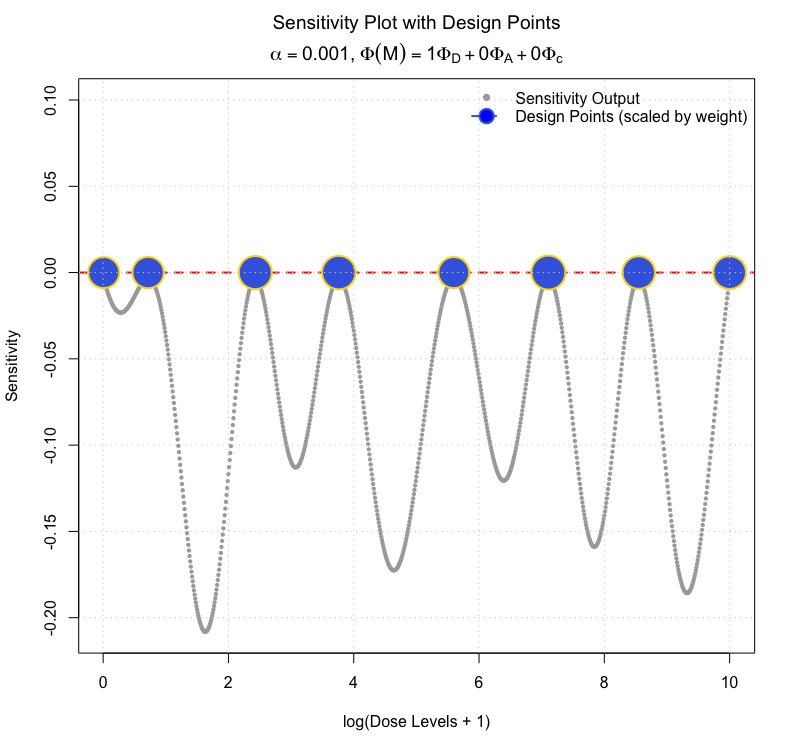}
        \includegraphics[width=0.3\textwidth]{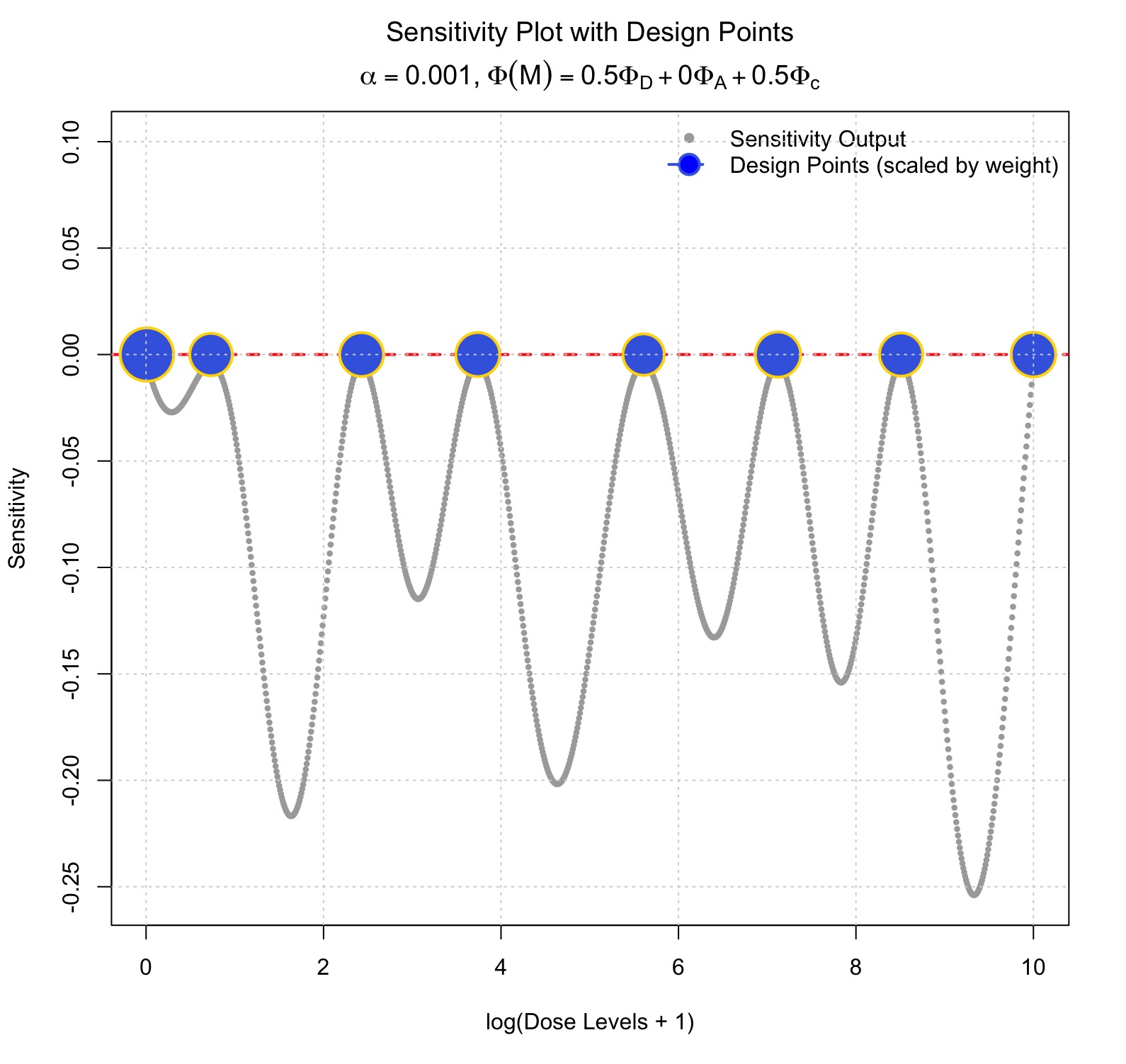}
        \includegraphics[width=0.3\textwidth]{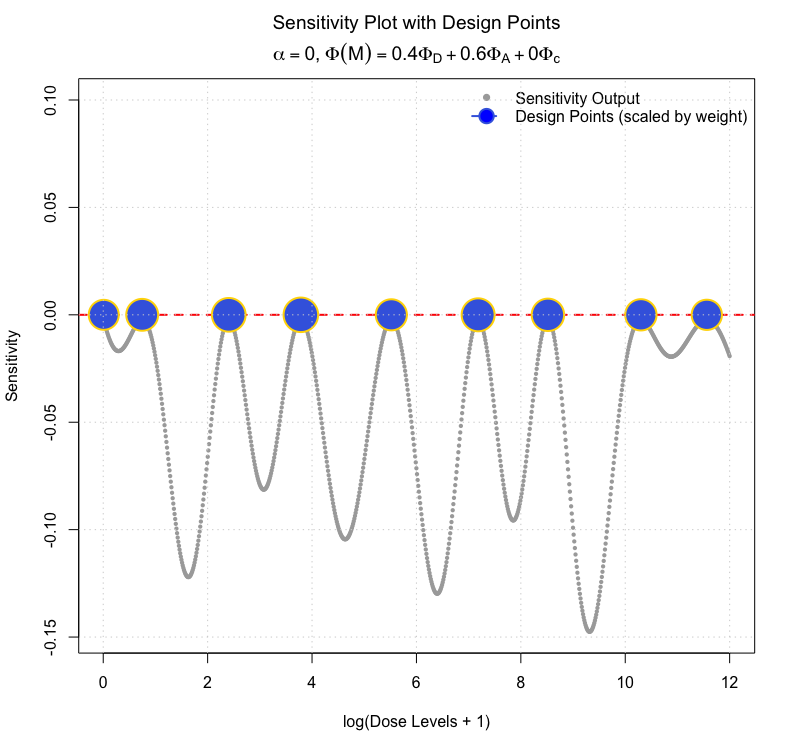}
        \caption{Sensitivity plots for the nine-parameter model.}
    \end{figure}

\end{document}